\documentclass[aps,prb,reprint,superscriptaddress,showkeys]{revtex4-1}

\usepackage{graphicx}
\usepackage{amssymb,amsfonts,amsmath}
\usepackage{bm}
\usepackage{changes}
\usepackage{hyperref}

\newcommand{\ceal}{CeAl$_{2}$}
\newcommand{\prni}{PrNi$_{5}$}

\newcommand{\cecual}{CeCuAl$_{3}$}
\newcommand{\ceaual}{CeAuAl$_{3}$}
\newcommand{\g}[1]{$|\Gamma_{#1}\rangle$}             
\newcommand{\gi}[2]{$|\Gamma_{#1}^{#2}\rangle$}       
\newcommand{\gp}[2]{$|\Gamma_{#1},{#2}\rangle$}       
\newcommand{\gip}[3]{$|\Gamma_{#1}^{#2},{#3}\rangle$} 
\newcommand{\gm}[1]{$|\widetilde{\Gamma}_{#1}\rangle$}        
\newcommand{\gim}[2]{$|\widetilde{\Gamma}_{#1}^{#2}\rangle$}  

\begin{document}

\preprint{APS/PRB}

\title{Magnetoelastic hybrid excitations in CeAuAl$_3$}

\author{Petr \v{C}erm\'{a}k} 
\affiliation{Charles University, Faculty of Mathematics and Physics, Department of Condensed Matter Physics, Ke Karlovu 5, 121 16, Praha, Czech Republic}
\affiliation{Forschungszentrum J{\"u}lich GmbH, J{\"u}lich Centre for Neutron Science at MLZ, Lichtenbergstr. 1, 85748 Garching,Germany}

\author{Astrid Schneidewind} 
\affiliation{Forschungszentrum J{\"u}lich GmbH, J{\"u}lich Centre for Neutron Science at MLZ, Lichtenbergstr. 1, 85748 Garching,Germany}

\author{Benqiong Liu}
\affiliation{Key Laboratory of Neutron Physics, Institute of Nuclear Physics and Chemistry, CAEP, Mianyang 621900, P. R. China}
\affiliation{Forschungszentrum J{\"u}lich GmbH, J{\"u}lich Centre for Neutron Science at MLZ, Lichtenbergstr. 1, 85748 Garching,Germany}

\author{Michael Marek Koza}
\affiliation{Institut Laue Langevin, 71 Avenue des Martyrs, 38042 Grenoble, France}

\author{Christian Franz}
\affiliation{Heinz Maier-Leibnitz Zentrum (MLZ), Technische Universit{\"a}t M{\"u}nchen, Lichtenbergstr. 1, 85748 Garching, Germany}
\affiliation{Physik-Department, Technische Universit{\"a}t M{\"u}nchen, 85748 Garching, Germany}

\author{Rudolf Sch{\"o}nmann}
\affiliation{Physik-Department, Technische Universit{\"a}t M{\"u}nchen, 85748 Garching, Germany}

\author{Oleg Sobolev}
\affiliation{Heinz Maier-Leibnitz Zentrum (MLZ), Technische Universit{\"a}t M{\"u}nchen, Lichtenbergstr. 1, 85748 Garching, Germany}
\affiliation{Institute for Physical Chemistry, Georg-August-University of G{\"o}ttingen, Tammannstr. 6, D-37077 G{\"o}ttingen, Germany}

\author{Christian Pfleiderer}
\affiliation{Physik-Department, Technische Universit{\"a}t M{\"u}nchen, 85748 Garching, Germany}


\date{\today}

\begin{abstract}
The interactions between elementary excitations such as phonons, plasmons, magnons, or particle-hole pairs, drive emergent functionalities and electronic instabilities such as multiferroic behaviour, anomalous thermoelectric properties, polar order, or superconductivity. Whereas various hybrid excitations have been studied extensively, the feed-back of prototypical elementary excitations on the crystal electric fields (CEF), defining the environment in which the elementary excitations arise, has been explored for very strong coupling only. We report high-resolution neutron spectroscopy and ab-initio phonon calculations of {\ceaual}, an archetypal fluctuating valence compound. The high resolution of our data allows us to quantify the energy scales of three coupling mechanisms between phonons, CEF-split localized 4f electron states, and conduction electrons. Although these interactions do not appear to be atypically strong for this class of materials, we resolve, for the first time, a profound renormalization of low-energy quasiparticle excitations on all levels. The key anomalies of the spectrum we observe comprise (1) the formation of a CEF-phonon bound state with a comparatively low density of acoustic phonons reminiscent of vibronic modes observed in other materials, where they require a pronounced abundance of optical phonons, (2) an anti-crossing of CEF states and acoustic phonons, and (3) a strong broadening of CEF states due to the hybridization with more itinerant excitations. The fact that all of these features are well resolved in {\ceaual} suggests that similar hybrid excitations should also be dominant in a large family of related materials. This promises a predictive understanding towards the discovery of new magneto-elastic functionalities and instabilities.
\end{abstract}

\maketitle

As the elementary excitations in solids reflect different aspects of the entire system of electrons, a wide range of coupling phenomena may be expected. For instance, the phonon-polariton, plasmon-polariton, or electro-magnon change their character continuously as a function of wavevector and energy \cite{Ashcroft}. In metallic systems well-defined fermion quasiparticle excitations have been reported, featuring electron-phonon interactions, as well as strongly dispersive paramagnon or polaronic dressing clouds \cite{1988:Lonzarich:JMMM}. The interest in such coupled elementary excitations has been driven by the search for and discovery of anomalous materials properties and electronic instabilities.  These may either be a direct consequence of the hybrid character of the low-energy dynamics, or result from from changes in lifetime and dispersion of pure elementary excitation, which affect the nature and range of their interactions.

Crystal electric fields (CEFs) may be considered one of the most important aspects in local-electron many-body physics, as they determine the environment in which elementary excitations arise. In this capacity, the CEFs control the nature and coupling of spin and orbital degrees of freedom, as well as electronic and magnetic anisotropies. The conventional static properties of the CEF excitations are extremely well understood and documented. In comparison, long-standing open questions concern the feed-back of elementary excitations on the CEFs, which leads to the formation of additional excitations beyond the expectations on the single ion level, as well as finite lifetimes and anomalous temperature dependences. Two primary mechanisms have been considered as the origin of such CEF properties. First, phonons may create CEF transitions between neighboring ions \cite{1978:sinha:handbook}, representing an important example of so-called magnetoelastic (ME) coupling \cite{1964:vlasov:jetp,2014:fennel:prl,2016:naji:jpcc,2017:boldyrev:prl}. Second, in metallic systems a coupling exists with particle-hole excitations \cite{BFK}. 

While various facets of the CEFs have been studied for nearly five decades, experimental evidence reflecting different coupling strengths as well as the full range of properties of the CEFs is surprisingly limited \cite{1979:steglich:jphyscol,2006:chapon:physb,2012:adroja:prl,2016:ruminy:prb,1988:loewenhaupt:jmmm,1991:fournier:prb}. This situation may be traced to the underlying interplay of energy scales, i.e., CEF excitations, phonons, particle-hole pairs, spin-orbit coupling, and magnetic interactions, all of which tend to be of similar strength, representing therefore a veritable chicken-and-egg-type of problem. 

Experimentally, the direct measurement of such complex low-energy dynamics requires a single crystal spectroscopic method with high resolution in both energy and momentum transfer. In contrast, as CEF excitations are traditionally assumed to be momentum-independent they have typically been determined in polycrystalline samples. Moreover, studies reporting momentum-resolved data on single crystals focus on a single facet of the hybrid character only, and lack the necessary comprehensive information.  In turn, there are a number of long-standing issues that may be resolved with targeted spectroscopic measurements, such as the feed-back of the ME coupling on the CEFs and spectrum of phonon excitations  \cite{1987:loewenhaupt:jmmm} the unambiguous identification of genuine hybrid excitations such as anticrossings \cite{1983:aksenov:physbc}, and the origin and nature of finite lifetimes and temperature dependences of the CEFs \cite{2006:blackburn:prb,1995:Bernhoeft:JPCM}. 

Taken together, these represent general problems, which are potentially relevant in any correlated-electron system. However, rare-earth intermetallics with moderate ME coupling are particularly suited for their study, because the different aspects of the hybrid character are well-defined and tractable. Namely, as both spin and orbital angular contributions generate the ME coupling, the well defined multiplett structure of the f-shells in rare earth compounds makes the ME coupling particularly tractable \cite{2018:mentink:arxiv}. 

The rich diversity of CEF-driven magnetoelastic phenomena in rare-earth based compounds has been recognized for a long time. Perhaps most prominent is the formation of a bound state between the CEF excitations and phonons in {\ceal}, known as vibronic-bound state (VBS) \cite{1979:loewenhaupt:prl, 2003:loewenhaupt:jpcm, 1982:thalmeier:prl,1984:thalmeier:jpc}. The key requirements of the VBS assumed so far are rather stringent, comprising a pristine (non-hybridized) CEF level at the same energy as a large phonon density of states. Moreover, as the phonons must have the same symmetry as the CEF level to permit hybridization and the phonon density of states must be excessive, it has been assumed that the VBS may only be formed with weakly dispersive (optical) phonons. Despite its conceptual importance, numerous questions are still unresolved in {\ceal}. These include the importance of phonons other than the optical modes, the observation of complex excitations at momentum transfers away from the $\Gamma$ point \cite{1971:marshall:monograph}, and the absence of Raman scattering of the highest level VBS \cite{1983:guntherodt:prl,1985:guntherodt:jmmm}. 

Several studies have proposed the formation of a VBS in other materials, considering an interplay of optical phonons with CEF excitations as in {\ceal}. For instance, consistent with this mechanism evidence for a VBS in polycrystalline PrNi$_2$ vanishes as the optical phonon levels shift to lower energies under doping \cite{1989:muhle:jmmm}. Putative evidence for vibronic states was also reported in cubic Ce$_{3}$Pt$_{23}$Si$_{11}$ \cite{2011:opagiste:prb} as well as the tetragonal systems CePd$_2$Al$_2$  \cite{2006:chapon:physb,2017:klicpera:prb} and {\cecual} \cite{2012:adroja:prl}. Moreover, vibronic excitations have even been proposed in rare-earth doped cuprates \cite{1996:ruf:physb}, as well as geometrically frustrated oxides such as Tb$_2$Ti$_2$O$_7$ \cite{2014:fennel:prl, 2016:ruminy:prb} and Ho$_2$Ti$_2$O$_7$ \cite{2018:gaudet:prb}, which underscores the wide-spread relevance. 

A different interplay with dispersive phonon branches as a function of momentum has been proposed in the regime of a steep crossing with the CEF excitations. For the case of dipolar magnetic interactions the emergence of an anticrossing in the presence of an applied magnetic field or spontaneous magnetic order has been observed, e.g., in Pr, PrAl$_2$, and TmVO$_4$ \cite{1991:thalmeier:handbook,1976:jensen:jpc,1979:houmann:prb,1976:purwins:aip,1975:thalmeier:zeitphysb,1975:kjems:prl}. On the other hand, an anticrossing in zero magnetic field may be expected theoretically between phonons and quadrupolar excitons, as putatively observed in PrAlO$_3$ \cite{1974:birgeneau:prb} and TbVO$_4$ \cite{1975:hutchings:jpc}. In intermetallic rare-earth compounds only indirect evidence for an anti-crossing has been reported for {\prni} \cite{1983:aksenov:physbc}, where a mere shifting of the exciton energies before and after the crossing could be detected, whereas the actual anticrossing could not be resolved. Interestingly, it has been speculated that the scattering of the phonons by magnetic excitons causes anomalous contributions to the thermal conductivity of {\prni} \cite{1988:reiffers:czechjpb}. 

Finally, the lifetime and temperature dependence of CEF excitations have been known to display pronounced deviations from the expected thermal population of single-ion states  \cite{1980:Lawrence:PRB,1985:Loewenhaupt:JMMM,2004:hense:jpcm,2006:blackburn:prb,2009:Iwasa:PRB,2009:Iwasa:JPCS}. In a seminal theoretical study Becker, Fulde and Keller (BFK) \cite{BFK} successfully explained the anomalous temperature dependences of the CEF occupation in terms of the temperature dependence of the interaction with particle-hole excitations in metallic systems. In turn, a large number of studies have attributed the broadening and temperature dependent effects observed experimentally to this mechanism. Of particular interest is the possibility of superconductive pairing due to CEF excitations and quadrupolar fluctuations as proposed in UPd$_2$Al$_3$ and PrOs$_4$Sb$_{12}$ \cite{2006:blackburn:prb,2009:Iwasa:PRB} \cite{2002:bauer:prb}. In this context, the broadening and thermal effects of the CEF are of central importance for the understanding of the metallic state and putative electronic instabilities.

\begin{figure}[t]
\centerline{
\includegraphics[width=0.8\linewidth]{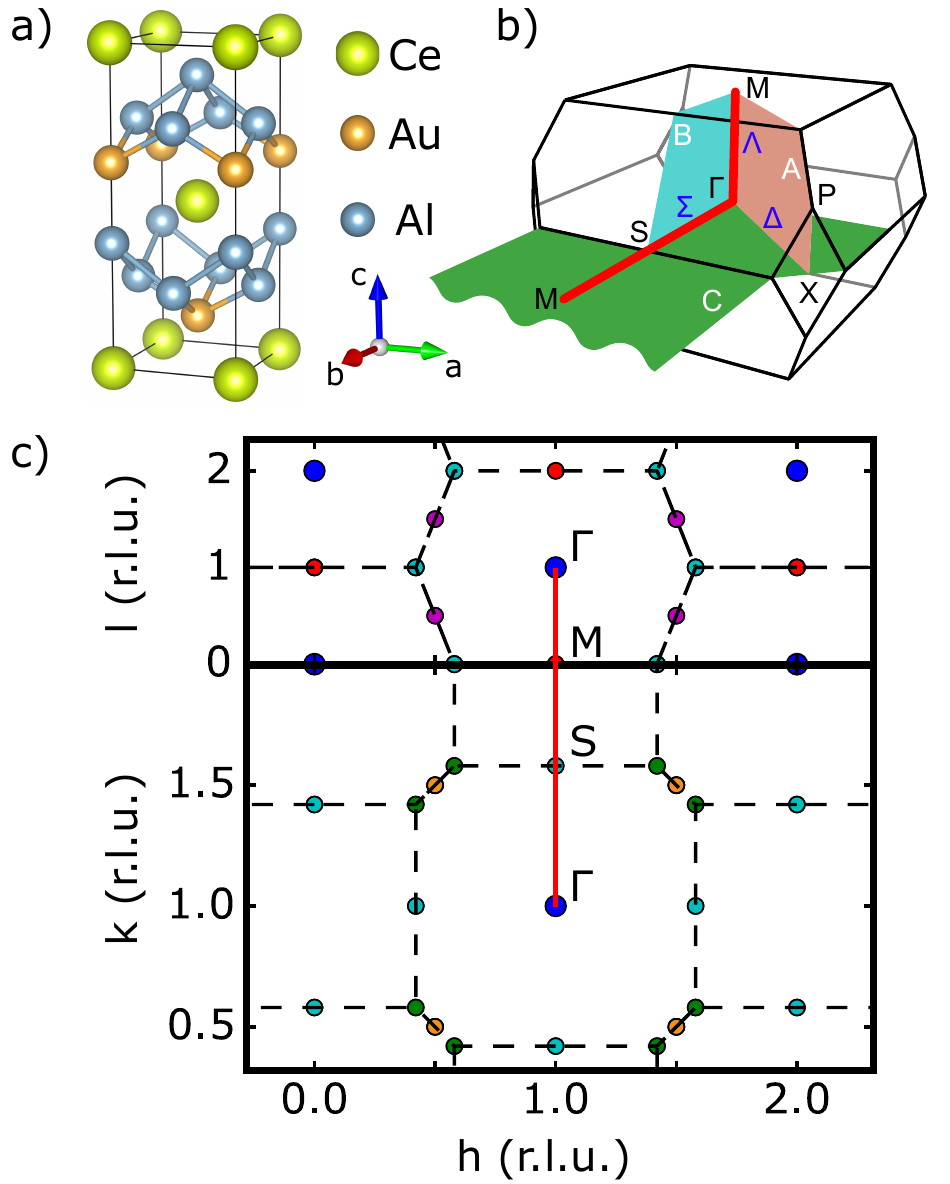}}
\caption{
Depiction of key characteristics of {\ceaual} in real and reciprocal space. (a) Crystallographic unit cell of {\ceaual}. The tetragonal BaNiSn$_3$ structure (space group $I4mm$, No. 107) lacks inversion symmetry \cite{1998:paschen:europhys,2014:klicpera:jmmm,2016:franz:jallcom}. (b) Brillouin zone of a body-centered tetragonal lattice (where $c>a$). High symmetry positions are marked according to the Bilbao notation \cite{Bilbao}, where points, lines and planes are denoted by black, blue and white letters, respectively. (c) ($h$, $k$, 0) and ($h$, 0, $l$) planes in reciprocal space. Locations at which data was recorded are marked by a red line.
}\label{figure2}
\end{figure}

In this paper we report a comprehensive inelastic neutron scattering study of the spectrum of low-lying excitations in {\ceaual} and ab initio phonon calculations. This compound is a member of the CeTAl$_3$ series (T=Ce, Au, Pd, Pt), which forms part of a wider family of BaAl$_4$-type materials \cite{2016:franz:jallcom}. Our findings for {\ceaual} are therefore directly relevant for a large number of systems featuring strong electronic correlations and magnetic order at the border of a quantum phase transition. Early measurements of the thermal, magnetic an charge transport properties of polycrystalline samples established that  {\ceaual} is a valence fluctuating compound with antiferromagnetic order below $T_\mathrm{N}=1.32\,$K \cite{1998:paschen:europhys}. 

The enhancement of the linear temperature dependence of the specific heat and quadratic temperature dependence of the resistivity ($\gamma=227$\,mJ\,mol$^{-1}$\,K$^{-2}$ and $A=5\,\mu\Omega$cmK$^{-2}$, respectively) are characteristic of a heavy Fermi liquid state. The CEF lifts the degeneracy of the Ce$^{3+}$ $J=5/2$ manifold, which has a characteristic impact on both the magnetic susceptibility and specific heat of the material. However, the first and second doublet at $T_{\rm I}=57\,{\rm K}$ and $T_{\rm II}=265\,{\rm K}$ are split from the ground state such that they have no bearing on the bulk properties and the enhancement of the Fermi liquid ground state.  Overall, {\ceaual} appears to be a typical Ce-intermetallic, which shares many of it physical properties with a large number of related materials. One unusual feature is its anisotropic reduced thermal conductivity (compared to non-magnetic LaAuAl$_3$). This has been interpreted in terms of enhanced magnetoelastic phonon scattering on the Ce ions taking into account the CEF splitting \cite{2000:aoki:prb}. 

The observation of a VBS state in CePd$_2$Al$_2$ \cite{2006:chapon:physb}, a related tetragonal compound, appears to be intimately related to a structural phase transition and underscores a strong interplay of the CEF excitations and phonons in this class of systems. Indeed, time of flight (TOF) neutron spectroscopy revealed also a VBS in polycrystalline {\cecual} \cite{2012:adroja:prl}, as confirmed recently in polarized single crystal neutron spectroscopy \cite{2018:klicpera:private} and in slightly off-stoichiometric samples \cite{2017:klicpera:inorgchem}. Here, too, electronic excitations are assumed to hybridize with optical phonons, which results in four doublets \gp{6}{0}, \gp{6}{1} and \gip{7}{1,2}{0}. This suggests that the symmetry of the lattice fluctuations imparts a different character to the VBS in tetragonal, as compared to cubic, systems. Yet, systematic time-of-flight neutron spectroscopy in polycrystalline CeRhGe$_3$ \cite{2012:adroja:prb} and {\ceaual} \cite{2015:adroja:prb} failed to detect a VBS. Moreover, the search for magnetoelastic phonon softening by inelastic x-ray scattering in {\cecual} and {\ceaual} has been inconclusive \cite{2017:tsutsui:physb}.

In contrast, revisiting the properties of single-crystal {\ceaual} in careful triple-axis neutron spectroscopy we find clear evidence for strong feed-back of phonons and particle-hole excitations on the CEF excitations. In particular we find a VBS, a clear anticrossing and substantial broadening of the CEFs. The observed behaviour is rather subtle and below the detection limit of neutron TOF spectroscopy in polycrystals \cite{2015:adroja:prb}, but consistent with the reduced thermal conductivity of {\ceaual} \cite{2000:aoki:prb}.  The rather weak coupling between CEFs, phonons and conduction electrons suggests that similar hybrid excitations must be generic in a wide range of materials.


\section*{Results}

The direct observation of the interplay between the lattice and electronic properties requires a spectroscopic probe which couples both to the lattice and electronic transitions with high resolution in both energy- and momentum transfer. Inelastic neutron scattering (INS) is uniquely suited to the task. Careful comparison of the INS signal in different Brillouin zones, and the temperature dependence of this response allows to distinguish nuclear and magnetic contributions to the spectral response, even without the need for polarization analysis. We have therefore used neutron triple axis spectroscopy at the spectrometers PANDA and PUMA at the Heinz-Maier Leibniz Zentrum at the Technical University of Munich. For technical details, we refer to the methods section and supplementary information \cite{SOM}. 

A high-quality and high-purity single crystals of {\ceaual} was grown by the optical floating zone method, using a ultra-high-vacuum compatible preparation chain  \cite{2016:bauer:scientificinstr}. The unit cell of the material (corresponding to the BaNiSn$_3$ structure type) is shown in Fig.\,\ref{figure2}\,(a). As previously reported, x-ray and neutron diffraction experiments revealed a very small amount of antisite disorder in these samples \cite{2014:klicpera:jmmm,2016:franz:jallcom}. This was also confirmed in recent NMR measurements in the CeTX$_3$ systems, which found minimal antisite disorder in {\ceaual}, compared to other members of the series \cite{2018:chlan:arxiv}.

Shown in Fig.\,\ref{figure2}\,(b) is the first Brillouin zone (BZ) of the body-centered tetragonal unit cell of {\ceaual}, where high symmetry directions are marked according to the Bilbao notation \cite{Bilbao}. Points, lines and planes in the BZ are denoted by black, blue and white letters, respectively. The relevant (h,k,0) and (h,0,l)-planes, and the trajectory of our measurement in reciprocal space are shown in Fig.\,\ref{figure2}\,(b) and (c) as indicated by a red line. Starting at the $\Gamma$ point in the centre of the Brillouin zone, this trajectory proceeded initially along the $c$-axis towards the zone boundary at the $M$ point. It then covered the properties in the $ab$-plane between the $M$ and the $S$ points. From here the trajectory returned back to the $\Gamma$ point. Since these directions of momentum transfer do not coincide with the main crystallographic directions of the primitive unit cell, neutrons couple to all polarizations of phonon modes. This proves to be very helpful in the discussion of our data presented below. 

\begin{figure*}[ht]
\centerline{
\includegraphics[width=0.95\linewidth, clip=]{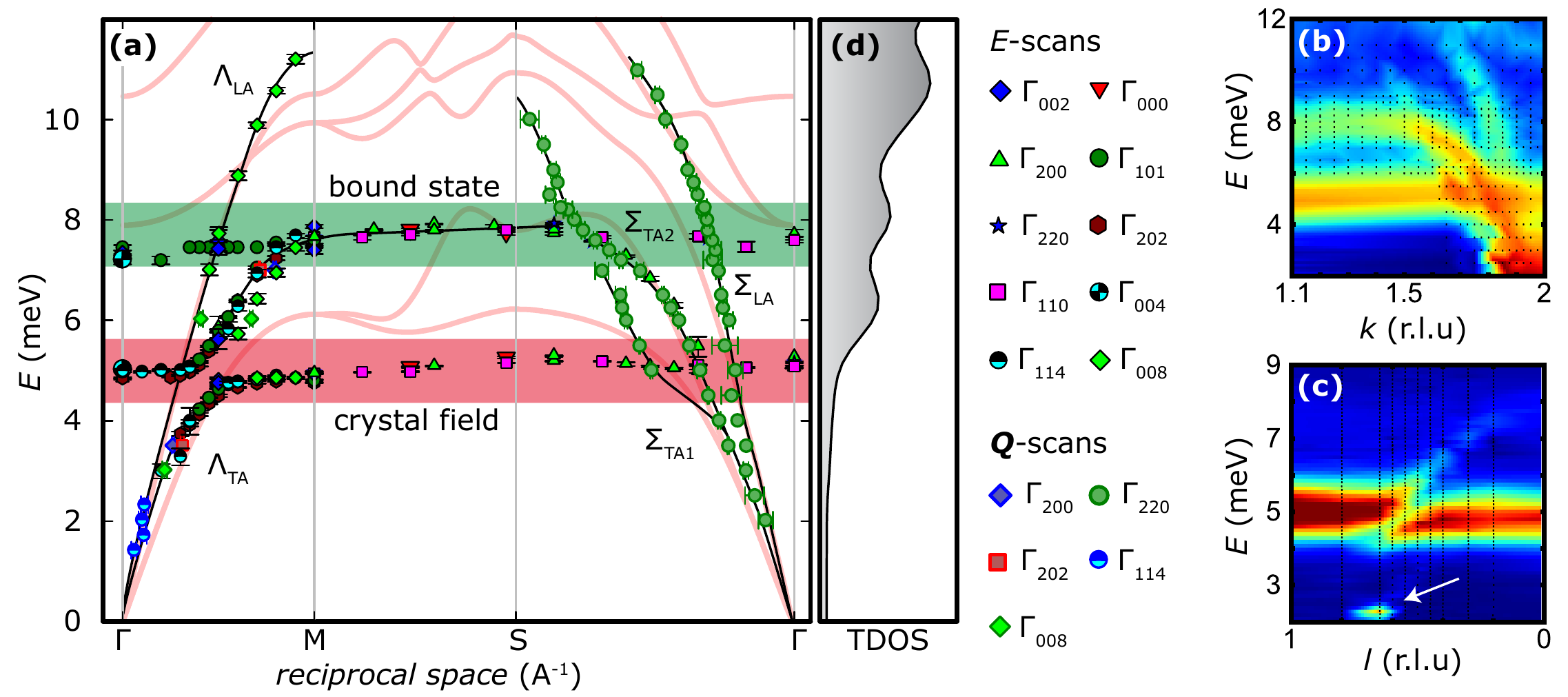}}
\caption{
Key characteristics of the neutron scattering excitation spectra of single crystal {\ceaual} observed in reciprocal space along $\Gamma - M - S - \Gamma$. (a) Energy versus reciprocal space map of {\ceaual}. Data points represent the peak positions of Gaussian fits (energy scans are denoted by a black border; momentum scans are denoted with a color-shaded border).  Blue, red and green shading denotes the Brillouin zone in which data were recorded (see also Fig.\,\ref{figure1}). The plots display the location of maxima only, but not the strength of the intensities. Red lines in background represent the results of ab-initio phonon calculations \cite{2018:liu:prb}. (b) Intensity map inferred from excitation spectra recorded between $M$ and $\Gamma$ along (2, $k$, 0). Above the crystal field excitations around 5\,meV a magneto-elastic hybrid excitation emerges around 8\,meV). (c) Intensity map inferred from excitation spectra recorded between $\Gamma$ and $M$ along (1, 0, $l$). A clear anti-crossing is observed. The feature marked by a white arrow represents spurious Bragg scattering.
(d) Calculated phonon density of states (cd. red lines in panel (a)). Maxima are observed at 6, 8 and 11\,meV, consistent with the bound state at $E_{\rm VBS}=7.9\,{\rm meV}$. The black line and color shading serve to guide the eye.
}\label{figure3}
\end{figure*}

An overview of the excitation spectra of {\ceaual} as a function of reduced scattering wave vector $\textbf{q}$ we observe is presented in Fig.\,\ref{figure3}\,(a). For any reduced scattering vector $q$, the spectra feature two flat excitations, marked by red and green shading. The flat excitations are crossed by strongly dispersive phonon modes branching out of the $\Gamma$ points. The interplay of these key characteristics results in two of the three main experimental observations of our study, notably (i) formation of a new bound state as marked by green shading, and (ii) distinct anti-crossing of acoustic phonons with the crystal field along the $\Gamma$-$M$ direction shown in Figs.\,\ref{figure3}\,(c). In addition we observe, (iii), a strongly enhanced broadening of the crystal field levels with temperature ((shown in more detail in Fig.\,\ref{figure4} and discussed below)). 

We now turn to a more detailed discussion of these observations. The crystal field excitation at $E_{\rm CF}=4.9\,{\rm meV}$ (red shading) may be attributed to the transition from the {\g6} ground state to the first excited doublet \gi{7}{1}, cf. Fig.\,\ref{figure3}\,(b). The energy of this transition is in excellent agreement with previous time-of-flight neutron spectroscopy and bulk data in a polycrystalline sample, which, however, do not allow to search for a momentum dependence \cite{2015:adroja:prb,1998:paschen:europhys}. The weak non-dispersive excitation at $E_{\rm VBS}=7.9\,{\rm meV}$ (marked in greed shading) is an unexpected new finding. This feature was not observed in the previous time-of-flight INS studies \cite{2015:adroja:prb}, probably due to the loss in spectral weight in the polycrystalline average.

The strongly dispersive excitations at the $\Gamma$-points may be clearly attributed to acoustic phonons as they emanate from nuclear Bragg peaks. Taking into account the tetragonal crystal symmetry, a longitudinal and a transverse acoustic branch are observed along the $\Gamma$ to $M$ direction, labeled as  $\Lambda_{\rm LA}$ and $\Lambda_{\rm TA}$, respectively. As illustrated in Fig.\,\ref{figure3}\,(c), these phonons display compelling evidence of an anticrossing with the crystal field excitation at $E_{\rm CF}$ in the three independent Brillouin zones studied, namely $(101)$, $(202)$ and $(114)$ \cite{SOM}. In contrast, for the $\Gamma$ to $S$ direction (labelled $\Sigma$) the two non-degenerate transverse acoustic phonons and a longitudinal acoustic phonon, denoted $\Sigma_{\rm TA1}$, $\Sigma_{\rm TA2}$ and $\Sigma_{\rm LA}$, respectively, cross the non-dispersive excitations at $E_{\rm CF}$ and $E_{\rm VBS}$ without apparent interaction (Fig. \ref{figure3}\,(b)). We did not detect any evidence of phonons in the vicinity of the zone boundary between the $M$ and $S$-points, probably because the intensity was too low or they coincided with the VBS. The intensity marked by the white arrow in Fig.\,\ref{figure3}\,(c) is a so-called Currat-Axe spurion, which is a common effect in triple-axis spectrometers. 
 
The momentum dependence of the intensity of the dispersion-less excitations at $E_{\rm CF}$ and $E_{\rm VBS}$ follows the from factor of the Ce$^{3+}$ ion (see Fig.~S6 in the Supplemental Material). This is strong evidence that these excitations are essentially of magnetic character. By contrast, the strongly dispersive excitations at the $\Gamma$-points are essentially due to nuclear scattering. Moreover, as a function of increasing temperature the intensity of both dispersionless excitations decreases strongly as shown in Fig.\,\ref{figure4}\,(a-c). This is qualitatively consistent with the thermal population of the first excited crystal field level, which provides further evidence for the magnetic character of these excitations. However, closer inspection reveals that the intensity decreases much faster than would be expected of a simple three-level system. We return to an accurate account of the temperature dependence below, which reveals dominant hybridization with the conduction electrons. 

As illustrated by the energy scans at the $\Gamma$ and M points shown in Fig.\,\ref{figure4}\,(d), the dispersionless excitation at $E_{\rm VBS}=7.9\,{\rm meV}$ strongly varies in intensity throughout the Brillouin zone. The intensity is large at the zone boundary and becomes very weak and difficult to discern at the zone centre.  This behaviour was observed in all Brillouin zones investigated, as listed in Fig.\,\ref{figure3}\,(a).  In stark contrast, no such variation was observed for the dispersionless excitation at $E_{\rm CF}=4.9\,{\rm meV}$, except near the $\Gamma$-point in the $(101)$ Brillouin zone (cf. Fig.\,S7 in the supplement). Taken together, this suggests that the formation of the dispersionless excitation at $E_{\rm VBS}=7.9\,{\rm meV}$, as well as the enhancement of the excitation at $E_{\rm CF}=4.9\,{\rm meV}$ at the $\Gamma$-point are driven by virtue of the coupling to the phonons. 


\section*{Discussion}   

For the discussion of our experimental results we assume that the magnetoelastic properties of {\ceaual} are dominated by the interactions of the 4$f$-electrons with the electrostatic crystal field. A Hamiltonian describing this situation is given by 
\begin{equation} \label{eq:1}
H=H_{\rm L}+H_{\rm CF}+H_{\rm ME},
\end{equation}
where $H_{\rm L}$ accounts for the kinetic and potential energy of the Bravais lattice, $H_{\rm CF}$ denotes the conventional crystal field Hamiltonian, and $H_{\rm ME}$ represents the single-ion magnetoelastic coupling of the spin to lattice strains. It is important to note that the phonons and crystal field excitations must share the same symmetry to couple directly. 
 
Treating the contribution of the Bravais lattice, $H_{\rm L}$, in a harmonic approximation we calculated the structural properties and spectrum of phonon excitations \cite{2018:liu:prb} using the Vienna Ab-initio Simulation Package (VASP) in the frozen-core projector augmented wave (PAW) method. The calculated lattice constants $a=4.335\,\text{\AA}$ and $c=10.844\,\text{\AA}$ were found to be in excellent agreement with experiment \cite{2016:franz:jallcom}. Further, the spectrum of phonon excitations was calculated using a finite difference method. The results are shown as faint red lines in Fig.\,\ref{figure3}\,(a) for energies up to $12\,{\rm meV}$ and also as total density of phonon states in Fig.\,\ref{figure3}\,(d). For the five atoms in the primitive unit cell of {\ceaual} the full phonon dispersion consists of 15 branches, which comprise three acoustic and twelve optical modes. Using group theoretical techniques that take into account all representations of the $C_{4\nu}$ point group are one-dimensional, an analysis of the irreducible representations for the high symmetry points establishes that the two acoustic branches along $\Gamma$ to $Z$ are degenerate. This justifies the assignment of the phonon branches in the vicinity of the $\Gamma$-points. At low energies the phonon branches are in excellent agreement with calculation. In contrast, for energies in the range 4\,meV to 9\,meV, we observe systematic deviations highlighting the presence of magneto-elastic coupling and the need for an advanced treatment.  

For the tetragonal symmetry of CeAuAl$_3$, point group $C_{4v}$ (\emph{4mm}), contributions of the Ce$^{3+}$ ions in the presence of a crystal electric field may be expressed as
\begin{equation}
H_{\rm CF}^0=B_2^0O_2^0+B_4^0O_4^0+B_4^4O_4^4,
\end{equation}
where $B_n^m$ and $O_n^m$ are the crystal electric field parameters and Steven's operators, respectively \cite{1989:balcar:book,1964:hutchings:ssp}. Due to the symmetry constraints all parameters should be real numbers \cite{2010:bauer:chin}.  In the paramagnetic phase the sixfold degenerate 4$f^1$ states of Ce$^{3+}$ ($J=\frac{5}{2}$) splits into three doublets. For our neutron data we find $B_2^0=1.203\,{\rm meV}$, $B_4^0=-0.001\,{\rm meV}$, and $B_4^4=\pm0.244\,{\rm meV}$. This corresponds to a {\g6} doublet ground state and two excited states, {\gi{7}{1}} and {\gi{7}{2}}, at energies of 4.95~meV and 24.3~meV, respectively. The associated eigenvectors are $|\Gamma_6\rangle=|\pm\frac{1}{2}\rangle$, $|\Gamma_7^1\rangle=-\alpha|\mp\frac{3}{2}\rangle+\beta|\pm\frac{5}{2}\rangle$, and $|\Gamma_7^2\rangle=\alpha|\mp\frac{5}{2}\rangle+\beta|\pm\frac{3}{2}\rangle$, with $\alpha=0.931$, $\beta=0.364$. These results are in excellent agreement with neutron time of flight spectroscopy of a powder sample of {\ceaual} \cite{2015:adroja:prb} as well as with the bulk properties \cite{1998:paschen:europhys}. Further details of the CEF analysis may be found in the Supplementary Information \cite{SOM}.
 
The calculations of the phonon spectrum and crystal field levels presented so far, clearly identify the dispersionless excitation at $E_{\rm VBS}=7.9\,{\rm meV}$ (green shading in Fig.\,\ref{figure3}\,(a)) as an additional state. It is tempting to interpret this intensity in terms of a vibronic state as observed in {\ceal} and {\cecual}. A comparison of the associated energy level scheme of {\ceal} and {\cecual} with {\ceaual} is shown in Fig.\,\ref{figure1}. For the cubic crystal structure of {\ceal}, the \g{7} doublet and \g{8} quartet of cerium in the CEF forms a set of new eigenstates.  By virtue of hybridization with optical phonons, this yields one electronic \gp{7}{0}, one phononic \gp{6}{1} doublet and two mixed \gim{8}{1,2} quartets, which are linear combinations of purely electronic and single-phonon states, depicted in Fig.\,\ref{figure1}(a).

Similarly, the formation of the VBS in the tetragonal crystal structure of {\cecual}, illustrated in Fig.\,\ref{figure1}\,(b), involves a hybridization of the \gi{7}{2} doublet with a high density of optical phonons to form a phononic \gp{6}{1} doublet and three mixed doublets, denoted \gm{6} and \gim{7}{1,2}. Thus, in both cases the formation of a Thalmeier-Fulde VBS presumes strong magnetoelastic interactions between weakly dispersive optical phonons and a nearby crystal field level. In contrast, in {\ceaual} at the energy of the putative VBS the phonon calculations do not yield the required high density of states. A low phonon density of states at at $E = 8\,\,{\rm meV}$, as well as $E \approx 6\,\,{\rm meV}$ and $E = 11-13\,\,{\rm meV}$ is also supported by our calculations (see Fig. \ref{figure3}\,(d) and neutron time-of-flight spectroscopy on polycrystalline samples of LaAuAl$_3$  \cite{2015:adroja:prb}. This means that one important prerequisite for the formation of a VBS is apparently absent.  

Nevertheless, our measurements suggest that a comparatively weak coupling between the excitation from the {\g6} to \gi{7}{1} state with acoustic phonons is sufficient for the formation of a VBS at $E_{\rm VBS}=7.9\,{\rm meV}$ as illustrated in Fig.\,\ref{figure1}\,(c). This implies that CEF excitations may form bound states with phonons that (i) do not cover large portions of the Brillouin zone, (ii) are dispersive and, (iii) of much reduced intensity. Further, with increasing magneto-elastic coupling the enhancement of the \gim{7}{1} level may be expected to shift down. In contrast, the newly created \gp{6}{1} level is shifting up. 

Using similar methods as reported in Refs.\,\cite{2012:adroja:prl,SOM}, the coupling constant may be roughly determined to be $g_{\rm VBS}=0.4\,{\rm meV}$ (for details of the calculations we refer to the Supplementary Information \cite{SOM}). The reduced intensity of the excitation at $E_{\rm VBS}=7.9\,{\rm meV}$ as compared with the excitation at $E_{\rm CF}=4.9\,{\rm meV}$ imply that the low phonon density of states is just sufficient to reach the threshold for the bound state to become measurable. In fact, it is interesting to speculate, whether the weak maximum of the calculated density of states at $E = 11\,\,{\rm meV}$ is just below the threshold, entailing an incipient VBS. 

\begin{figure}[ht]
\centerline{
\includegraphics[width=1.0\linewidth]{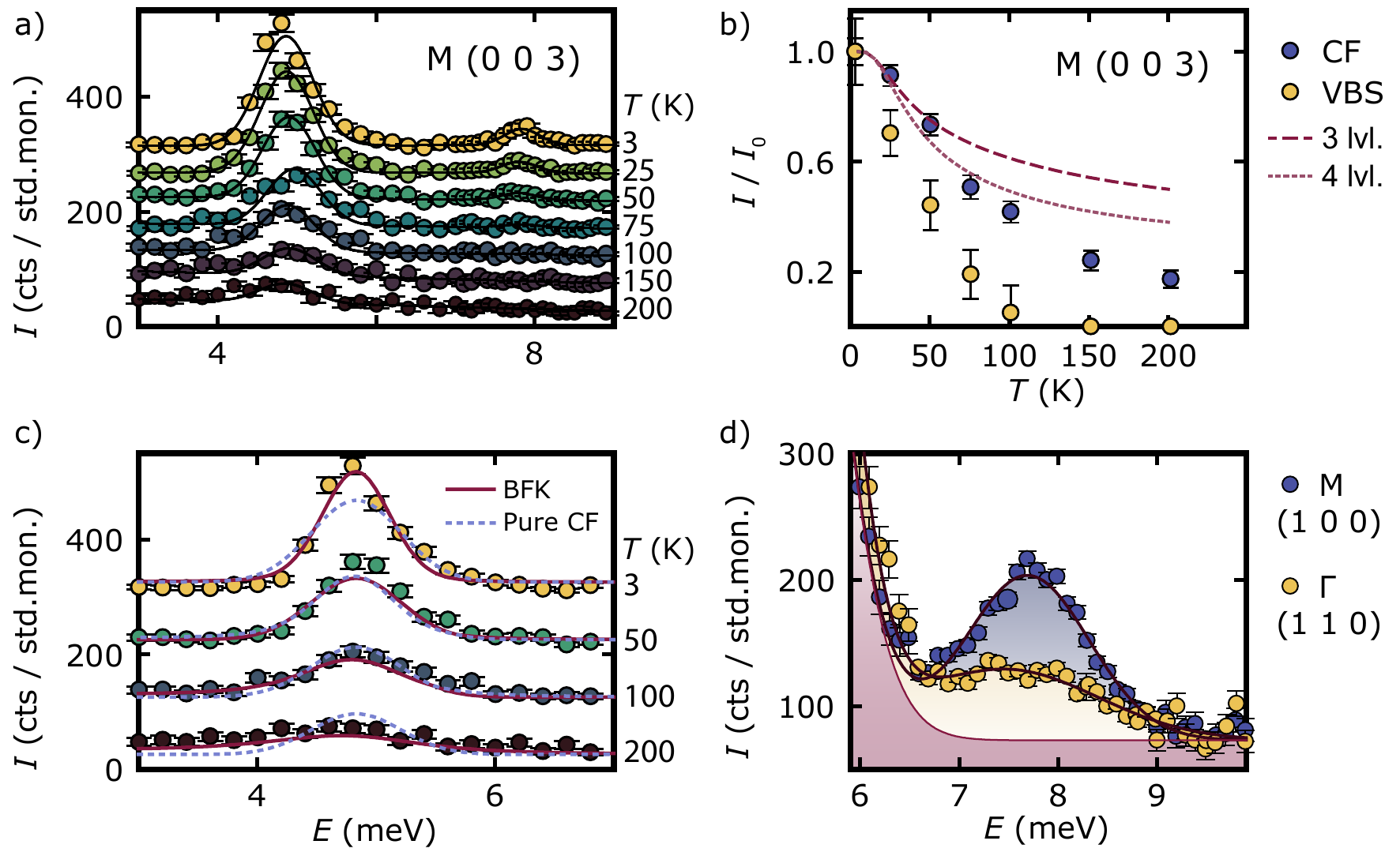}}
\caption{Temperature dependence of the CEF and vibronic bound state at $E_{\rm CF}=4.9\,{\rm meV}$ and $E_{\rm VBS}=7.9\,{\rm meV}$, respectively. 
(a) Energy scans at the $M$ point at $(0 0 3)$ at various temperatures (cf. values on right hand side). Scans are shifted vertically by 60 cts for clarity. Solid lines are the result of fitting of two independent Gaussians. 
(b) Integrated intensity of the CEF and vibronic bound state. Intensities are normalized to 3 K. Red dashed lines show the calculated temperature dependence of the excitations at the $M$ point for a three- and a four-level system. 
(c) Selected scans shown in panel (a) as fitted with a pure crystal field model with fixed intensities in a 3-level model (dotted lines) and a model based on the Becker-Fulde-Keller model \cite{BFK} (solid lines).
(d) Energy scans through the vibronic excitation at the $\Gamma$- and the $M$-point, i.e., at the zone-centre and at the border of the BZ, respectively. Dark solid lines represent a Gaussian fit to the data. Red shading denotes the background and the pure crystal field excitation.
}\label{figure4}
\end{figure}

The magneto-elastic interactions may also be expected to affect the spectrum of phonon and crystal field excitations where their interplay is strongest. This is indeed the case in the regime of the anticrossing \cite{SOM,2018:liu:prb}. Indeed, the anticrossing also proves to be a direct consequence of the magnetoelastic coupling. We follow considerations first reported for PrNi$_5$  \cite{1983:aksenov:physbc}, where the Hamiltonian of the magnetoelastic coupling, $H_{\rm ME}^I$, describes direct coupling between the deformations of the lattice and the 4$f$ shell. In a group theoretical analysis of this expression \cite{1965:callen:pr} , the energy of the associated coupled quadrupole-phonon excitation can be stated as 
\begin{equation}
\omega_{q\pm}^2=\frac{E^2+\omega_{0}^2}{2} \pm \sqrt{\left(\frac{E^2-\omega_{0}^2}{2}\right)^2 + 16 \alpha^2 E \omega_0^2 g_{\rm AC}} 
\label{eq-anticrossing}
\end{equation}  
where $\omega_{q\pm}$ represent the energies of the two anticrossing excitations; $\omega_{0}$ is the phonon energy which depends on $\bm{k}$, $E$ is the non-dispersive energy of the crystal field level involved and $g_{\rm AC}$ is an effective coupling constant related to the renormalization of elastic constant, see supplement for details \cite{SOM}. A fit of our data yields $g_{\rm AC}$ = 12.1(2)\,$\mu$eV. 

\begin{figure}[ht]
\centerline{ 
\includegraphics[width=1.0\linewidth]{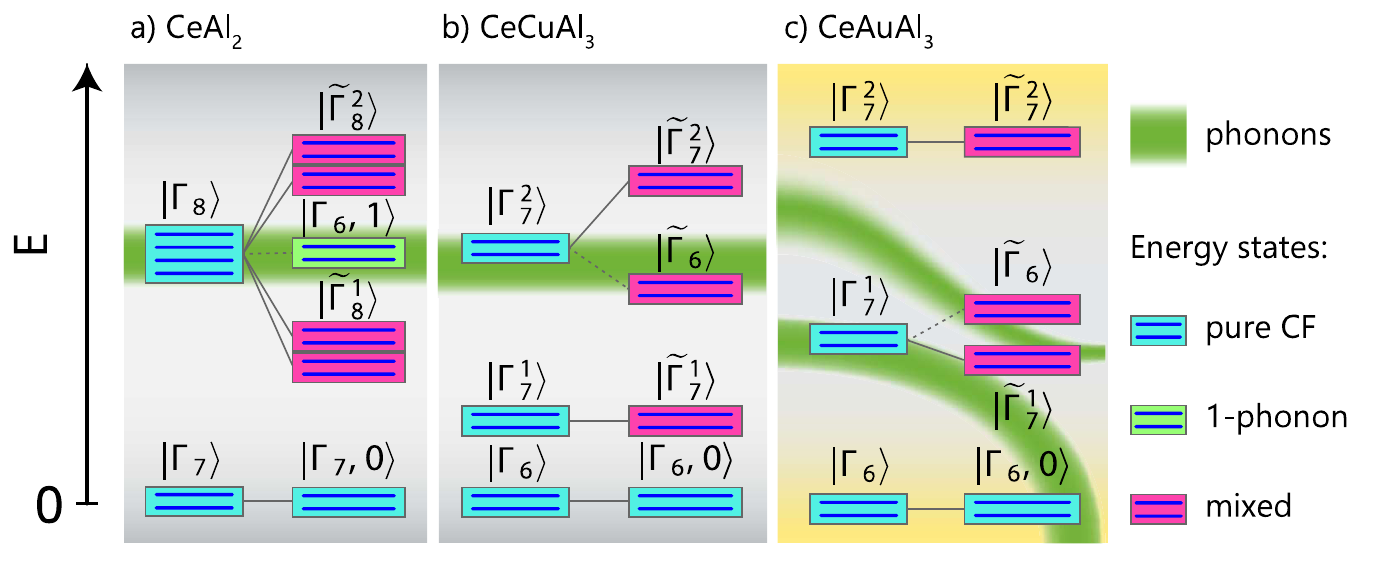}}
\caption{Comparison of CEF - phonon excitations and emergence of magnetoelastic bound states in selected $f$-electron materials. (a) Vibron in {\ceal}. Under the action of the three magneto-elastic operators $\mathcal{O}_{1,2,3}$ the unperturbed \g{8} splits into 5 doublets and a dispersionless vibron forms the CEF coincides with an optical phonon \cite{1979:loewenhaupt:prl,1982:thalmeier:prl,1985:fulde:advances}. (b) Magneto-elastic hybrid excitation in {\cecual}. Under the magneto-elastic operator ${\mathcal{O}_2^2}$ the unperturbed doublet \gp{7}{1} splits into 2 doublets and a bound crystal field-phonon mode has been reported \cite{2012:adroja:prl}. (c) Three magneto-elastic characteristic in {\ceaual}, notably a bound crystal field-phonon mode, anti-crossing, and crystal-field damping due to conduction electrons. The phonon density of states (green shading) illustrates the importance of dispersive phonon modes.
}\label{figure1}
\end{figure}

The magneto-elastic coupling is also reflected in the temperature dependence of the scattering intensity, shown in terms of energy scans at the $M$ point in Fig.\,\ref{figure4}\,(a). It is instructive to consider the reduced scattering intensity, $I/I_0$, normalized to its value at low temperatures as shown in Fig.\,\ref{figure3}\,(b) for $E_{\rm CF} = 4.9\,{\rm meV}$ and $E_{\rm VBS} = 7.9\,{\rm meV}$. At 200\,K we observe a large reduction of the intensity of the crystal field level at $E_{\rm CF}$ by $\sim80\,\%$, whereas the intensity of the excitation at $E_{\rm VBS}$ already vanishes above 100\,K. In contrast, a reduction of only 50\,\% would be expected of the intensity at $E_{\rm CF}$ for 200\,K, when thermally populating the three crystal field excitations determined in the standard analysis, which ignores the weak mode at $E_{\rm VBS}$. This situation improves slightly with a reduction of 60\,\% at 200\,K for four crystal field levels when additionally taking into account the mode at $E_{\rm VBS}$. However, the agreement is still far from satisfactory. 

When additionally considering the coupling to the conduction electrons following the suggestion of Ref.\,\cite{2017:tsutsui:physb}, a BFK model of crystal field line broadening \cite{BFK} provides excellent agreement with our data.The fitting procedure incorporates the code of Keller \cite{McPhase}, where technical details may be found in the Supplementary Information \cite{SOM}. As shown in Fig.\,\ref{figure4}\,(c) the improved account of the peak intensity in the BFK model is also reflected in an improved account of the energy dependence. The associated unitless coupling constant, $g_{\rm BFK} = 0.022(0)$, is remarkably small. While the BFK model already provides a satisfactory agreement with the broadening further improvements may be expected when taking into account the interactions with the spectrum of phonons. A full analysis of these contributions lies beyond the present capabilities of established computational techniques \cite{BFK-phonon}. 

As summarized in Tab.\,\ref{sampletable}, the coupling constants $g_{\rm BFK}$ and $g_{\rm VBS}$  in {\cecual} are smaller than in {\ceal}. This highlights that the spectrum of low-lying CEF excitations may be modified profoundly, even for systems with rather weak magneto-elastic coupling.


\section*{Conclusions}

In summary, we find a remarkable combination of different facets of the interactions of phonons and conduction electrons with the CEF excitations in {\ceaual}. The magneto-elastic hybrid character of the excitation comprises the formation of a bound state akin to the VBS in CeAl$_2$, a well-resolved anticrossing, and strong damping of the CEF levels. The combined observation of all of these different effects in the same material, the importance of acoustic phonons in their formation, and the comparatively weak coupling imply that these effects must be present in many materials.  This underscores, that moderately interacting rare-earth compounds offer tractable insights towards a predictive understanding of magneto-elastic functionalities and instabilities.

\begin{table}
\caption{Coupling constants as compared with the literature: (i) vibronic bound state, $g_{\rm VBS}$, (ii) anticrossing, $g_{\rm AC}$, and (iii) damping through conduction electrons, $g_{\rm BFK}$.}
\label{sampletable}
\begin{minipage}{1.0\linewidth}
\begin{tabular}{lllll}
compound	  &	$g_{\rm VBS}$  [$\mu$eV]    	& $g_{\rm AC}$  [$\mu$eV] & $g_{\rm BFK}$ [unitless]    &  Ref.\\
\hline
{\ceaual} 	  & $\approx$ 400   				& 12.1(2)			  & 0.022(0) 	& this study \\
{\ceal}   	  & 540 						&                  			  & 0.06            	& \cite{1984:thalmeier:jpc} \\
{\cecual} 	  & 800  						&                  			  &                 	& \cite{2012:adroja:prl}    \\
{\prni}	  &             					& $<$ 4	\footnote{This value was obtained by fitting the data reported in Ref. \cite{1983:aksenov:physbc}. See supplement \cite{SOM} and Fig. S6 for details.}	  &  			& \cite{1983:aksenov:physbc}  \\
\end{tabular}
\end{minipage}
\end{table}

\section*{Methods}

\subsection*{Materials preparation and quality}
A high quality single crystal of {\ceaual} was prepared by optical float-zoning under ultra-high vacuum compatible conditions \cite{2016:bauer:scientificinstr}. For the inelastic neutron scattering experiments a crystal with a mass of approximately 2\,g was used. High sample purity was confirmed by means of resistivity, magnetisation and specific heat of small pieces cut from the same ingot \cite{2014:franz:thesis,2016:franz:jallcom}. The correct BaNiSn$_3$-type structure and high crystalline quality were confirmed by powder and Laue x-ray diffraction as well as neutron diffraction \cite{2016:franz:jallcom}. Great care was exercised to confirm the correct crystal structure \cite{2016:franz:jallcom}. Neutron diffraction established that anti-site disorder is negligible in the present samples \cite{2016:franz:jallcom}.

\subsection*{Neutron scattering}
Inelastic neutron scattering measurements were carried out on the triple axis spectrometers PUMA and PANDA at MLZ, Garching \cite{PUMA, PANDA}. For details of the experimental setup and the momentum and energy ranges covered in our experiments we refer to the supplementary information \cite{SOM}. The sample was cooled with a pulse-tube cooler. Data were recorded at a temperature of 5~K unless stated otherwise, i.e., well above the antiferromagnetic transition temperature at 1.3\,K. 

\subsection*{Theoretical calculations}
Ab initio calculations have been carried using VASP and the frozen-core projector augmented wave (PAW) method. Taking into account weak interactions of the rare-earth (RE) ions with phonons in terms of magnetoelastic effects, an analytical expression for the hybridization of quadrupole excitations and phonons from the poles of the one-phonon Green-function is derived. For details see \cite{SOM} and Ref.\cite{2018:liu:prb}.

\begin{acknowledgments}
We acknowledge discussions with P. Thalmeier, W. Wulfhekel, J. Kulda, M. Loewenhaupt, M. Janoschek, M. Wilde and M. Klicpera. BL is supported by the Nat. Nat. Science Found. of China (Grant No. 11305150), and Direct. Found. of China Acad. of Eng. Phys. (Grant No. YZ2015009). The work of PC was supported by the Czech Science Found. (Grant No. 17-04925J). CP and AS acknowledge support through DFG TRR80 and DFG grant WI3320-3. CP acknowledges support through ERC AdG ExQuiSid. 
\end{acknowledgments}

\bibliography{bibliography}

\end{document}


\preprint{APS/PRB}

\title{Supplemental Material for: Magnetoelastic hybrid excitations in non-centrosymmetric CeAuAl$_3$}
\author{P. \v{C}erm\'{a}k, A. Schneidewind, B. Liu, M. M. Koza, C. Franz, R. Sch{\"o}nmann, O. Sobolev, C. Pfleiderer}
\noaffiliation

\newcommand{\ceal}{CeAl$_{2}$ }
\newcommand{\ceptsi}{CePt$_{3}$Si }
\newcommand{\cecual}{CeCuAl$_{3}$ }
\newcommand{\ceaual}{CeAuAl$_{3}$ }
\newcommand{\g}[1]{$|\Gamma_{#1}\rangle$}
\newcommand{\gi}[2]{$|\Gamma_{#1}^{#2}\rangle$}
\newcommand{\gp}[2]{$|\Gamma_{#1},{#2}\rangle$}
\newcommand{\gip}[3]{$|\Gamma_{#1}^{#2},{#3}\rangle$}

\maketitle

\section*{Experimental details}
\label{exp}
\begin{figure}[ht]
\centering
 \includegraphics[width=0.6\linewidth]{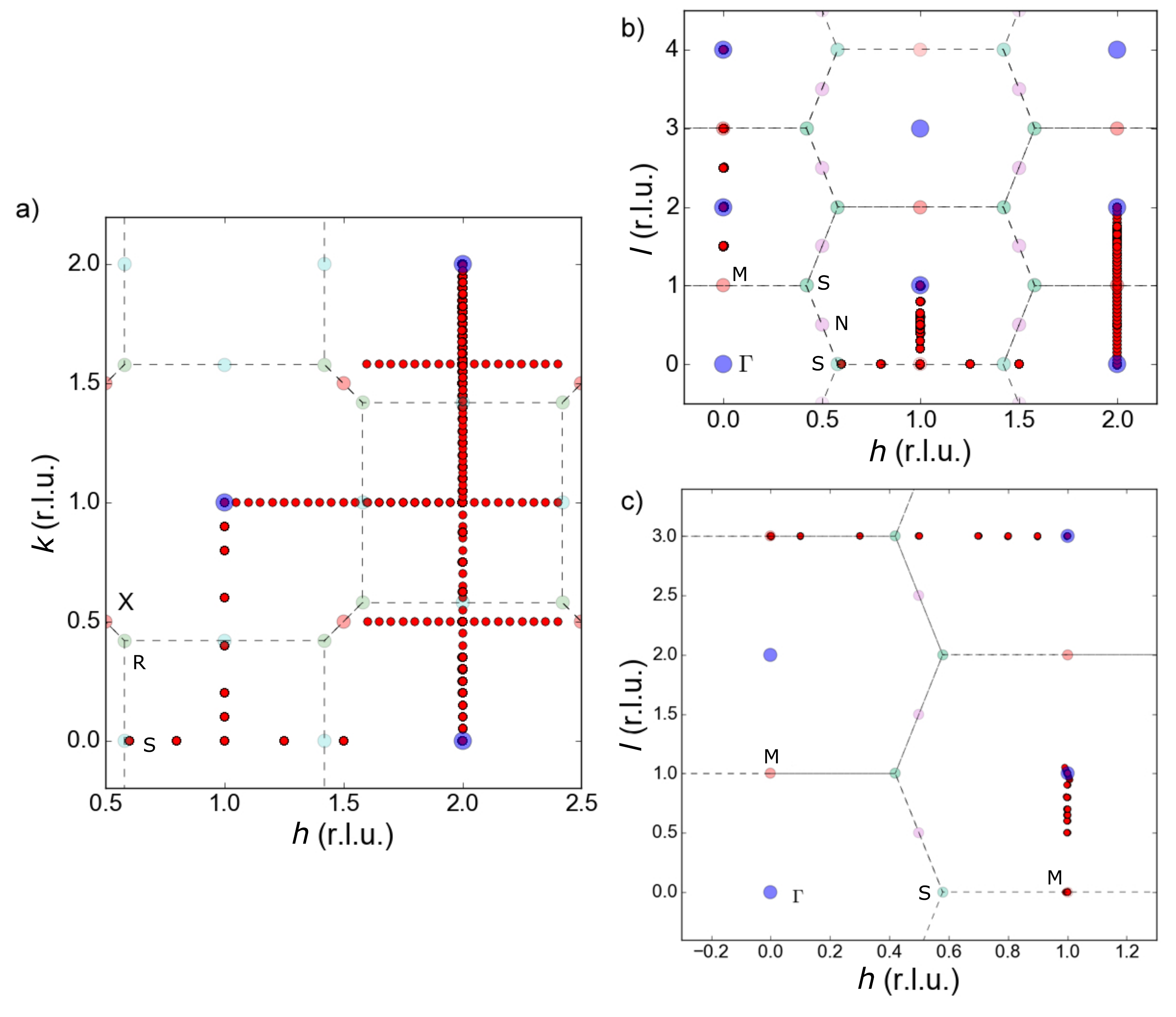}%
 \caption{\label{rs-all} Location of all data points measured on PUMA in the ($h$, 0, $l$) sample orientation (panel (a)), and the ($h$, $k$, 0) sample orientation (panel (b)).
(c) Projection of the location of all data points measured at PANDA.}
\end{figure}

The neutron scattering measurements were carried out on the triple-axis spectrometers PUMA and PANDA at MLZ \cite{PANDA, PUMA}. The experimental setups involved double-focussing PG002 monochromators and analyzers and a closed-cycle cryostat for sample cooling. In the measurements at PUMA, data were recorded in the ($h$, 0, $l$) and ($h$, $k$, 0) plane using constant $k_{\rm f} = 2.662 \text{ \AA}^{-1}$ and two PG-filters after the sample. To determine the resolution, selected scans were recorded at the PG004 reflection of the analyzer. At PANDA, high-resolution measurements were carried out in ($h$, 0, $l$) orientation of the sample using $k_{\rm f} = 1.57 \text{ \AA}^{-1}$ with cooled Be-filter. Further data were recorded using $k_{\rm f} = 1.97 \text{ \AA}^{-1}$ with a PG-filter after the sample. The positions in $q$-space where spectra were collected are shown in Fig. \ref{rs-all}.

\section*{Crystal field analysis}
\label{CF}

The pure crystal field levels without taking into account perturbations were analyzed using the standard Stevens formalism \cite{1964:hutchings:ssp}. The position of the Ce atoms in \ceaual is characterized by the point group symmetry $C_{4v}$ (\emph{4mm} in int. notation). This allows to reduce the general crystal field Hamiltonian to:
\begin{equation}
  H_{CEF}^0=B_2^0O_2^0+B_4^0O_4^0+B_4^4O_4^4,
  \label{eq-purecf}
\end{equation}
where $B_n^m$ and $O_n^m$ are the CEF parameters and Steven's operators, respectively \cite{1989:balcar:book,1964:hutchings:ssp}. In the paramagnetic phase the sixfold degenerate Ce$^{3+}$ state ($J=\frac{5}{2}$), which corresponds to a 4$f^1$ configuration, is split into three doublets:
\begin{equation}
  \begin{split}
    |\Gamma_6\rangle &= |\pm \frac{1}{2}\rangle \\
    |\Gamma_7^{(1)}\rangle &= \alpha| \pm \frac{5}{2}\rangle - \beta| \mp \frac{3}{2} \rangle \\
    |\Gamma_7^{(2)}\rangle &= \beta| \pm \frac{5}{2}\rangle + \alpha| \mp \frac{3}{2} \rangle 
  \end{split}
\end{equation}

Neutron scattering of the crystal field transitions between these states on a single crystal sample bears the advantage, that different ground states may be distinguished directly from a comparison of the different crystallographic directions. The intensity of the transition between two CF levels depends on the matrix elements of the transition matrix
\begin{equation}
  S(\bm{Q}, E)_{E_n \rightarrow E_m} = c \cdot \big( f(\bm{Q}) \big)^{2}
  P_n \sum_\alpha{\left|\langle\Gamma_m|J^{\alpha}|\Gamma_n\rangle\right|^2} \delta(E - E_n - E_m),
\end{equation}
where $f(\bm{Q})$ is the magnetic form factor, $P_n$ expresses thermal occupation of level $E_n$, $\alpha$ = $x$, $y$ or $z$, and $\bm{J}$ is the total angular momentum operator. Since neutrons are sensitive to the magnetic moment perpendicular to $\bm{Q}$ only, it is possible to observe the different $J^{\alpha}$ matrices.

Because of the simple form of the $J^z$ matrix, it does not contribute to transitions from/to the \g6 state. Consequently for the case of the \g6 ground state, the $J^x$ and $J^y$ components can be measured together along the crystallographic $c$ direction whereas only $J^x$, $J^y$ is measured along the $b$, $a$ direction respectivelly. In addition, transition intensities $J^x$  and $J^y$ are the same for the \g6 ground state in tetragonal structure. On the other hand, for the \gi{7}{(1)} or \gi{7}{(2)} ground states there is not such simplification.

\begin{figure}[ht]
\centering
   \includegraphics[width=0.6\linewidth]{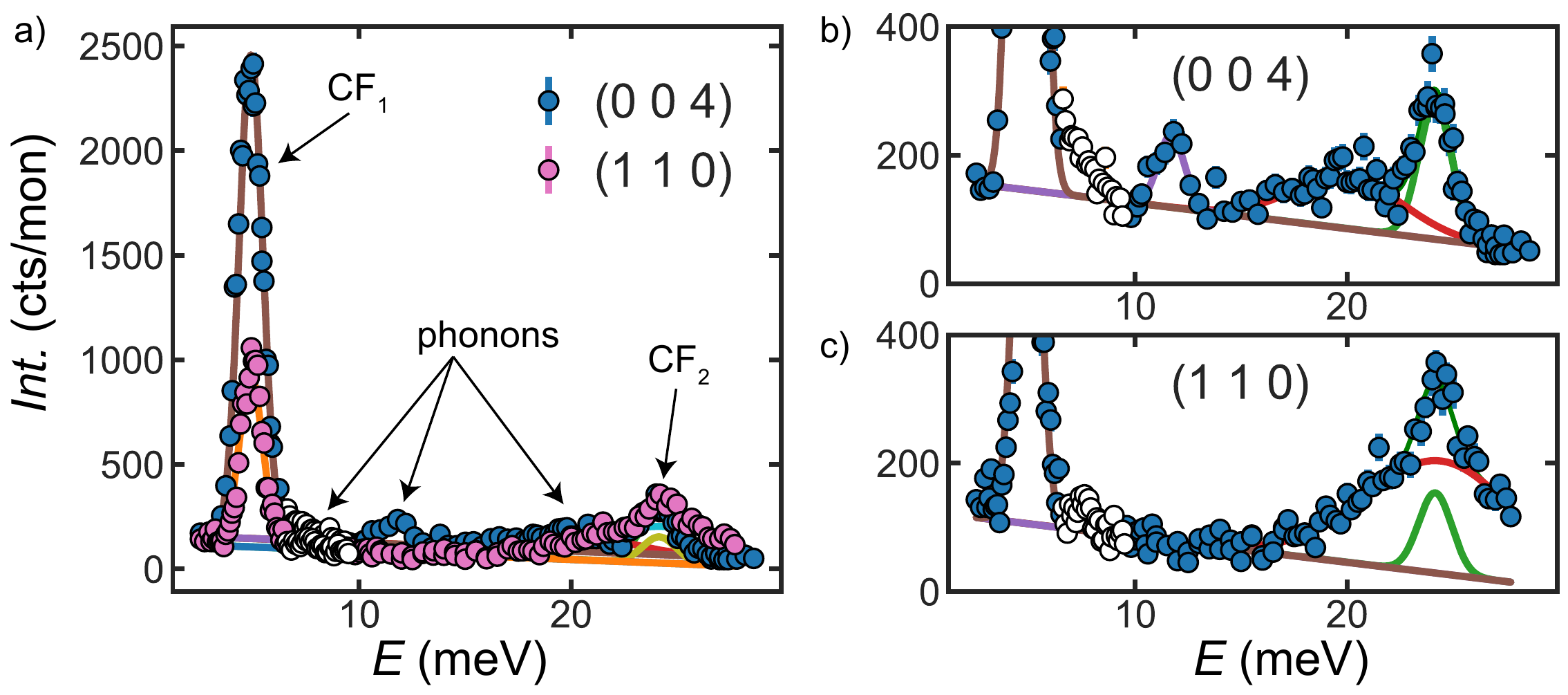}%
   \caption{\label{cf-fits} (a) Energy scans in two different crystallographic directions. The monitoring rate of the data was corrected for higher order contamination. Arrows indicate the observed crystal field levels as well as contamination by phonons. Data marked by white symbols were not taken into account in the analysis as the contamination due to phonons was difficult to fit. (b) and (c) Close-up view of the data. The lines represent Gaussians.
   }
 \end{figure}

Energy scans were recorded at $Q$ = (0, 0, 4) and $Q$ = (1, 1, 0) between 2 and 28 meV on the spectrometer PUMA. An analytic correction of the monitor for higher order scattering was applied to the raw data. Both spectra are shown in Fig. \ref{cf-fits}(a). The  intensities of the peaks inferred from fits of the data are summarized in table \ref{table-cf-fits}. As stated above, for the case of the \g6 ground state we find $I_{(004)} \approx 2 \cdot I_{(110)}$, otherwise $I_{(004)} < 2 \cdot I_{(110)}$. Thus, the ground state is in excellent agreement with  \g6, consistent with Ref.\,\cite{2015:adroja:prb}.

\begin{figure}[ht]
\centering
   \includegraphics[width=0.6\linewidth]{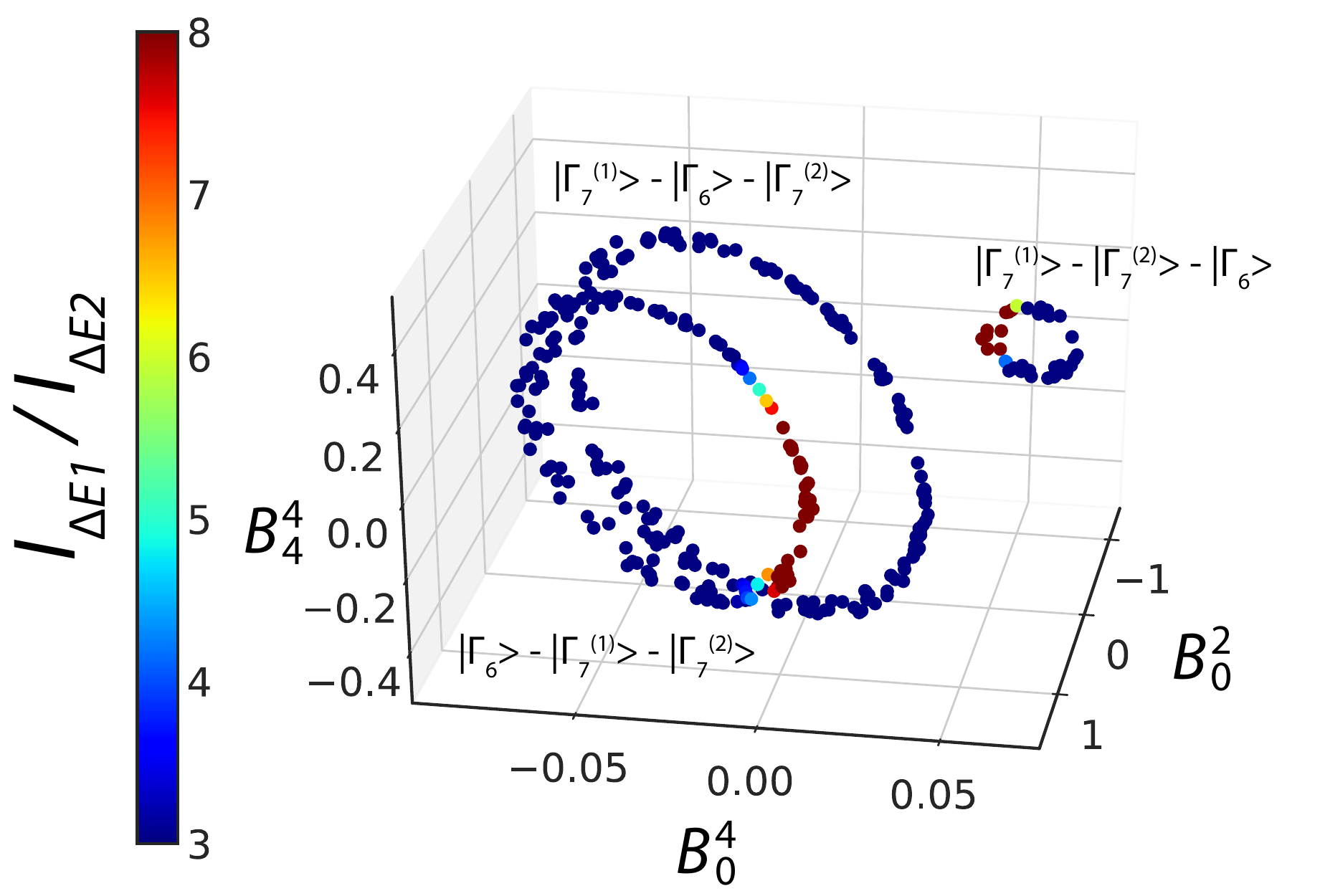}%
   \caption{\label{fig-cf-3D} Visualization of the possible crystal field parameters resulting in an energy splitting of
   $\Delta_1 = 4.885 \textrm{meV} \pm 5\%$ and $\Delta_2 = 24.16 \textrm{meV} \pm 5\%$.
   Different rings represent a different order of CEF states. 
   The color coding serves to highlight the ratio between the intensities of two crystal field peaks.
   }
 \end{figure}

For the detailed determination of the crystal field parameters,  Monte-Carlo calculations were used to visualize the possible solutions in a depiction of $B_2^0$, $B_4^0$ and $B_4^4$ shown in Fig. \ref{fig-cf-3D}. This analysis underscores, that for given ratio of the CF excitation to the ground state, there exist only two independent solutions (and another two with opposite sign of $B_4^4$ parameter which is indistinguishable using neutron scattering). Two of these four solutions are in agreement with the \g6 ground state. A Generalized Reduced Gradient (GRG2) \cite{1978:lasdon:acm} nonlinear optimization algorithm was used to refine the detailed $B$ parameters, resulting in $B_2^0 = 1.203, B_4^0 = -0.00095$ and $B_4^4 = \pm 0.244$~meV. The parameters accounting for the crystal field splitting calculated from this set of $B$ values are denoted in Table \ref{table-cf-fits}.

 \begin{table}[ht] 
  \centering
  \caption{Intensities of the CF transitions}  \label{table-cf-fits}
  \begin{tabular}{l|lllll|lll}
         & \multicolumn{5}{c}{measured}                           &  \multicolumn{3}{c}{calculated}  \\
         $Q$      &	$I_{CF1}$    & $I_{CF2}$   & $I_{CF1}$/$I_{CF2}$   & $E_1$ & $E_2$  &  $I_{CF1}$/$I_{CF2}$ & $E_1$ & $E_2$ \\
      \hline
   (0, 0, 4) & $3220(50)$ & $420(80)$ &  & \multirow{3}{*}{ $4.885(7)$ } & \multirow{3}{*}{ $24.16(9)$ }  &  \multirow{3}{*}{ $6.6$ }  & \multirow{3}{*}{ $ 4.949 $ } & \multirow{3}{*}{ $ 24.26 $ }  \\
   (1, 1, 0) & $1140(30)$ & $250(70)$ &  & & & &  \\ 
   averaged $J_x$ & $1450(30)$ & $ 220(50) $ & 	$6.6(1.7) $ & &  \\
  \end{tabular}
\end{table}

\section*{Becker-Fulde-Keller theory}
\label{BFK}

\begin{figure}[ht]
\centering
   \includegraphics[width=0.6\linewidth]{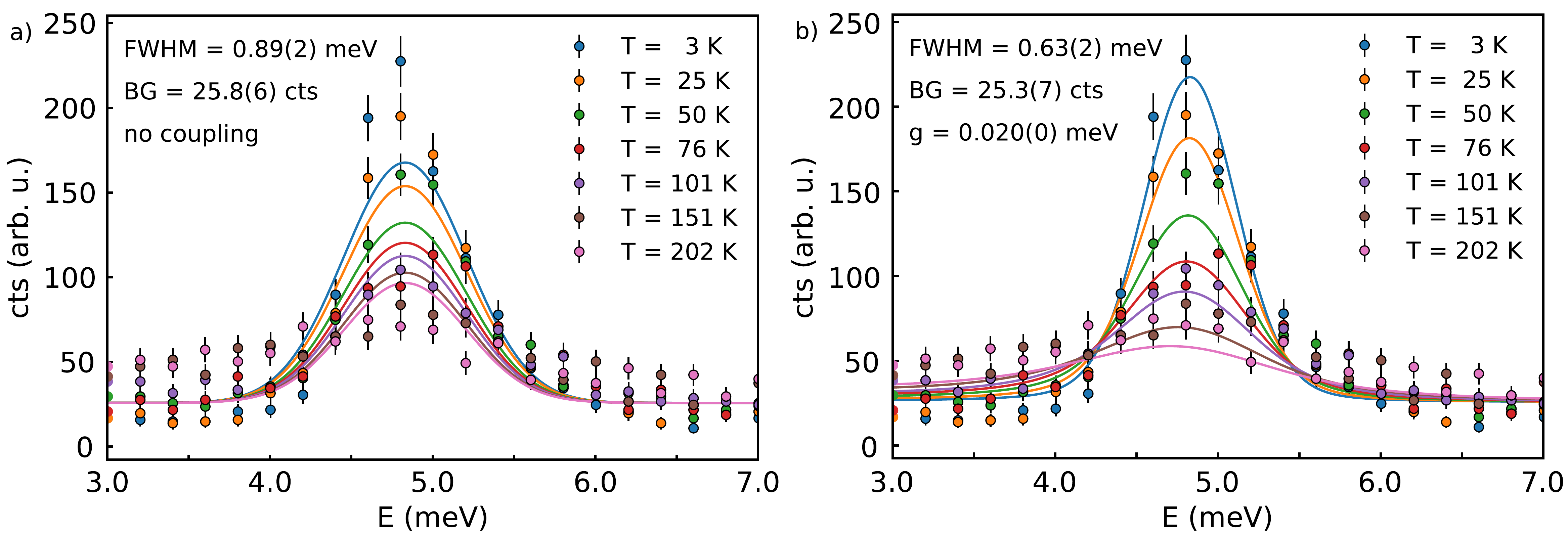}%
   \caption{\label{fig-bfk} Energy scans at Q = (0, 0, 3) at different temperatures
   (see als legend for details). 
   a) Data were fitted with a Gaussian of the same width and fixed intensity ratios
   regarding population of pure CF excitation from the ground state.
   b) Data were fitted with Becker-Fulde-Keller theory. 
   Resulting fit parameters are stated in the figure.
   }
 \end{figure}

Pure crystal field excitations are expected to have negligible lifetime. In turn, the width of the CEF peaks is expected to be limited by the resolution of the spectrometer and temperature independent. As shown in Fig. \ref{fig-bfk}(a) this is not the case for CeAuAl$_3$. Data shown here were simultaneously fitted at seven different temperatures using the same constant background and same width of the peaks. An alternative analysis is reported in the main text, where the same data were fitted with an independent sloping background and an independent width of the peaks (cf. Fig. 3(a) in the manuscript). While the quality of the fits is very good, the temperature dependence of the intensities inferred from the fits does not follow the expected behaviour as shown in Fig. 3(b) in the main text.

The reason is that part of the magnetic intensity transfers into the background with increasing temperature. This effect is well described by a theoretical model proposed by by Becker, Fulde and Keller (BFK) \cite{BFK}. In order to compare the experimental data with the predictions of BFK theory, the original code written by Keller turned available as open source as a part of the McPhase suite \cite{McPhase} was used. Unfortunately, the original code in FORTRAN is very slow and not suitable for fitting. The code was therefore enhanced and recompiled with a PYTHON interface and published as a part of CrysFiPy software \cite{crysfipy}. The results of the fit are shown in Fig. \ref{fig-bfk}(b), where all seven energy scans were fitted simultaneously with the BFK model (using a single FWHM and a flat background) and a coupling constant $g_{\rm BFK}$ = 0.022 was obtained.

\section*{Analysis of the anticrossing}
\label{AC}

In order to account for the anticrossing and to evaluate the coupling energy, the results of Aksenov and coworkers for PrAl$_2$ \cite{1981:aksenov:jourphysf} and later PrNi$_5$ \cite{1983:aksenov:physbc} were adapted. A detailed procedure may be found in Ref.\,\cite{2018:liu:prb}. In the following the main steps of the calculations are summarized. A double-time Green function method was used to describe the coupling of the lattice to magnetic ions for cubic symmetry and zero field. While in PrNi$_5$ the anticrossing due to  the dipole transition $\Gamma_4\ - \Gamma_5$ is nicely observed by neutron scattering, in case of PrAl$_2$ the quadrupolar $\Gamma_3\ - \Gamma_4$ would not be neutron active without coupling to the lattice. For CeAuAl$_3$, the calculation needs to be extended to tetragonal symmetry. To do so the $f$-electrons are treated as strictly localized, because the first CEF excitation at $E$=59.2\,K \cite{2012:adroja:prl} lies well above $T_\textrm{K}$ \cite{1998:paschen:europhys}. For a standard Hamiltonian of RE metallic compounds, taking the magnetoelastic interaction into account, the one-phonon Green-function (GF) in the random phase approximation may be expressed as \cite{1981:aksenov:jourphysf,1983:aksenov:physbc}
\begin{equation}
  \label{eq-green}
D(\bm{q},\omega)=\Big[\Big(D^0(\bm{q},\omega)\Big)^{-1}-\sum_{mn}{G_{mn}(\bm{q},\omega)}\Big]^{-1},
\end{equation}
where $D^0(\bm{q},\omega)$ is the phonon GF in the harmonic approximation, $\bm{q}$ is wave vector of phonons, and
\begin{equation}
  \label{eq-quadexcitation}
G_{mn}(\bm{q},\omega)=\frac{E_{mn}V_{mn}^1V_{nm}^1(f_m-f_n)^2}{\omega^2-E_{mn}^2},
\end{equation}
where $f_m=\exp{(-\beta E_m)}/\sum_m{\exp{(-\beta E_m)}}$ is the occupation number of one-ion states, and $E_{mn}$ are energies of transitions between CEF levels. $V_{mn}^1=\langle m|V_1(\bm{J},\bm{q})|n\rangle$ are matrix elements of the operators $V_1(\bm{J},\bm{q})$ which are functions of the $\bm{J}$ operators depending on the symmetry of the magnetoelastic coupling \cite{1975:dohm:zeitphysb}.

The total Hamiltonian of the system may be written as
\begin{eqnarray}
H_{CEF}^f=H_{CEF}^0+H_{me}^I,
\end{eqnarray}
$H_{CEF}^f$ denotes the interaction of the spin $\bm{J}$ of the $f$-th ion with the crystal field caused by the other ions, with the zeroth-order term $H_{CEF}^0$ being the conventional crystal field Hamiltoninan described by Eq.\,(\ref{eq-purecf}) and also solved there. $H_{me}^I$ is the one-ion magnetoelastic interaction, coupling the spin system to the strain. At low temperature, only the lowest-lying level is occupied. In the following the low-energy transition $|\Gamma_6\rangle\to|\Gamma_7^1\rangle$ at $E_{\Gamma_6\Gamma_7^1}=4.9$\,meV is considered, which crosses the acoustic phonon.

The one-ion magnetoelastic Hamiltonian, accounts for the direct coupling between the deformations of the lattice and the 4$f$ shell. This Hamiltonian may be constructed according to group theory \cite{1965:callen:pr}. Let $\mathcal{S}_i^{\Gamma,j} (i=1,2,\cdots,n)$ be the spin functions which form a basis for the $n$-dimensional representation $\Gamma$, different sets being distinguished by $j=$1, 2. Then the one-ion contributions $H_{me}^I$ to the magnetoelastic Hamiltonian may be written as
\begin{equation}
  \label{eq-mel-hamiltonian}
H_{me}^I=-\sum_{\Gamma}{\sum_{j,j'}{\widetilde{B}_{jj'}^\Gamma(f)\sum_i{\epsilon_i^{\Gamma,j}\mathcal{S}_i^{\Gamma,j'}(f)}}},
\end{equation}
where $\widetilde{B}_{jj'}^\Gamma$ is a phenomenological magnetoelastic coupling constant, and $\epsilon_i^{\Gamma,j}$ represents linear combinations of the first-order strain components $\epsilon_{xx}$, $\epsilon_{yy}$, $\epsilon_{zz}$, $\epsilon_{xy}$, $\epsilon_{yz}$, and $\epsilon_{zx}$. The shear strains are defined as $\epsilon_{xy}=\frac{1}{2}(\frac{\partial u_y}{\partial x}+\frac{\partial u_x}{\partial y})$. 

It is reasonable to assume that the crystal is constrained so that all "antisymmetric strains" such as $\frac{1}{2}(\frac{\partial u_y}{\partial x}-\frac{\partial u_x}{\partial y})$, which correspond to homogeneous rotations of the crystal, vanish. Whereas in zero applied magnetic field the rotational magnetoelastic interaction may be expected to contribute to the phonon dispersion curves for finite $\bm{q}$-values \cite{1979:fulde:handbook}, the coupling is expected to give corrections of a few percent only. Therefore, these terms will not be considered in this work. $\mathcal{S}^{\Gamma,j'}(f)$ corresponds to the spin operators of the ion at site $f$, as shown in Table \ref{table-strain}. As Eq.\,(\ref{eq-mel-hamiltonian}) has been derived assuming homogeneous strains, it is only valid for long-wavelength acoustic phonons \cite{1991:thalmeier:handbook}.

\begin{table*}[ht]
\centering
\caption{Strain functions and one-ion spin operators for the tetragonal system.}
\label{table-strain}
\begin{tabular}{ll}
\hline
Strain functions $\epsilon_i^{\Gamma,j}$&   One-ion operators $\mathcal{S}^{\Gamma,j'}(f)$\\
\hline
$\epsilon^{\alpha 1}\equiv\frac{\sqrt{3}}{3}(\epsilon_{xx}+\epsilon_{yy}+\epsilon_{zz})$ &$1$\\
$\epsilon^{\alpha 2}=\sqrt{\frac{2}{3}}[\epsilon_{zz}-\frac{1}{2}(\epsilon_{xx}+\epsilon_{yy})]$ &$3J_z^2-J(J+1)$\\
$\epsilon^\gamma=\frac{\sqrt{2}}{2}[\epsilon_{xx}-\epsilon_{yy}]$ & $J_x^2-J_y^2=\frac{1}{2}(J_+^2+J_-^2)$\\
$\epsilon^\delta=\sqrt{2}\epsilon_{xy}$ &  $P_{xy}=\frac{1}{2}(J_xJ_y+J_yJ_x)$\\
$\epsilon_1^\epsilon=\sqrt{2}\epsilon_{yz}$ &  $P_{yz}=\frac{1}{2}(J_yJ_z+J_zJ_y)$\\
$\epsilon_2^\epsilon=\sqrt{2}\epsilon_{zx}$ &  $P_{zx}=\frac{1}{2}(J_xJ_z+J_zJ_x)$
\\
\hline
\end{tabular}
\end{table*}

Thus the one-ion magnetoelastic Hamiltonian for the tetragonal symmetry (point group $4mm$) may be expressed directly as \cite{1990:morin:book}
\begin{eqnarray}
H_{me}^I(f)=&-&\widetilde{B}^{\alpha_1}\epsilon^{\alpha 1}\Big[3J_z^2-J(J+1)\Big]\nonumber\\
&-&\widetilde{B}^{\alpha_2}\epsilon^{\alpha 2}\Big[3J_z^2-J(J+1)\Big]\nonumber\\
&-&\widetilde{B}^\gamma\frac{\sqrt{2}}{2}(\epsilon_{xx}-\epsilon_{yy})(J_x^2-J_y^2)\nonumber\\
&-&\widetilde{B}^\delta\frac{\sqrt{2}}{2}\epsilon_{xy}(J_xJ_y+J_yJ_x)\nonumber\\
&-&\widetilde{B}^\epsilon\frac{\sqrt{2}}{2}\epsilon_{yz}(J_yJ_z+J_zJ_y)\nonumber\\
&-&\widetilde{B}^\epsilon\frac{\sqrt{2}}{2}\epsilon_{xz}(J_xJ_z+J_zJ_x),
\end{eqnarray}

In other words, the magnetoelastic Hamiltonian may be considered as the strain derivative of the crystal field Hamiltonian. The two-ion magnetoelasticity is in turn related to the modification of the two-ion magnetic interactions by the strain.

The one-ion magnetoelastic Hamiltonian is sufficiently general when considering the main effects. Therefore, in this study two-ion magnetoelastic interactions were ignored which may lead to structural and magnetic phase transitions \cite{1975:dohm:zeitphysb}. In addition, the experimental data (even under an applied magnetic field) was analysed in terms of $V_1(\bm{J},\bm{q})$, i.e., neglecting the second order magnetoelastic interaction as well as the linear rotational interaction.

We specialize our example of magneto-elastic interaction of shear deformations within the 
$x-z$ plane (all $\frac{\partial u_\alpha}{\partial\beta}=0$ except $\frac{\partial u_x}{\partial z}$ and $\frac{\partial u_z}{\partial x}$). 
Then we have
\begin{equation}
V_1(\bm{J},\bm{q})\sim e_z(q_x)q_x(J_xJ_z+J_zJ_x),
\end{equation}
where $\bm{e}(\bm{q})$ is the polarization vector for the phonon.

For the $\Gamma-$M direction, only the matrix elements determined by $\Gamma_6^A-\Gamma_7^{1B}$ and $\Gamma_6^B-\Gamma_7^{1A}$ differ from zero. Using the wave functions of one-ion states, the matrix elements for Eq.\,(\ref{eq-quadexcitation}) may be obtained in the form
\begin{equation}
|V_{\Gamma_6^A\Gamma_7^{1B}}^1|=|V_{\Gamma_6^B\Gamma_7^{1A}}^1|=2\sqrt{2} \alpha \widetilde{B}^\epsilon |\bm{q}|,
\end{equation}
where $\widetilde{B}^\epsilon$ is the magnetoelastic coupling constant and $|\bm{q}|$ is reciprocal wavevector. In order to compare the coupling in different materials, we introduce an effective coupling constant $g_{\rm AC}$, which is directly related to the renormalization of elastic constant (see Eq. 37 in Ref.\,\cite{1991:thalmeier:handbook} with $g_{\rm AC} = \widetilde{g}^2$):
\begin{equation}
g_{\rm AC} = \frac{(\widetilde{B}^\epsilon)^2}{c_{44} \Omega} = \frac{(\widetilde{B}^\epsilon)^2}{v_s^2 m_c} = \frac{\hbar^2 (\widetilde{B}^\epsilon)^2 q^2}{\omega_0^2 m_c},
\end{equation}
where $c_{44}$ is an elastic constant, $\Omega$ is the volume of the primitive cell, $v_s$ is the speed of sound and $m_c$ is the mass of the primitive cell.
Using Eqs.\,(\ref{eq-green}) and (\ref{eq-quadexcitation}), the coupled quadrupole-phonon dispersion $\omega(\bm{q})$ may be determined by the poles of the Green-Function (\ref{eq-green}), i.e., by setting the denominator of Eq.\,(\ref{eq-quadexcitation}) equal to zero \cite{1975:dohm:zeitphysb}:
\begin{eqnarray}
  (\omega_\mathbf{q}^2)^2-\omega_\mathbf{q}^2E_{\Gamma_6\Gamma_7^1}^2-\omega_0^2\omega_\mathbf{q}^2+\omega_0^2E_{\Gamma_6\Gamma_7^1}^2
  -\frac{\hbar^2}{m_c}E_{\Gamma_6\Gamma_7^1}(|V_{\Gamma_6^A\Gamma_7^{1B}}|^2+|V_{\Gamma_6^B\Gamma_7^{1A}}|^2)=0.
\end{eqnarray}

\begin{figure}[ht]
\centering
   \includegraphics[width=0.6\linewidth]{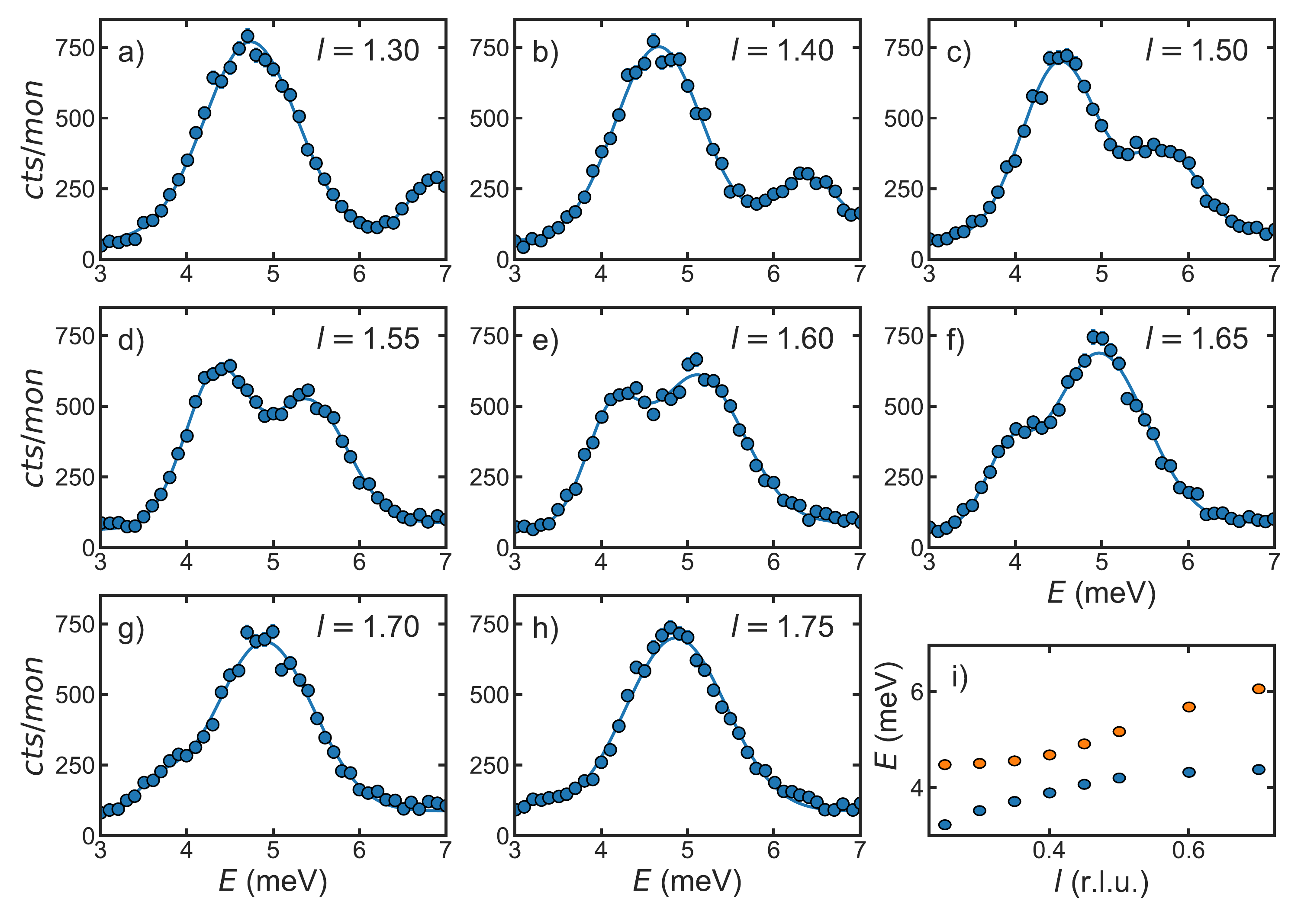}%
   \caption{\label{ac-all-fits} (a-h) Energy scans recorded at the PUMA spectrometer at positions (20$l$). Data are fitted with two Gaussians and the same sloping background.
   (i) Dispersion relations inferred from the maximum of the fits.}
  \end{figure} 

\begin{figure}[ht]
\centering
      \includegraphics[width=0.6\linewidth]{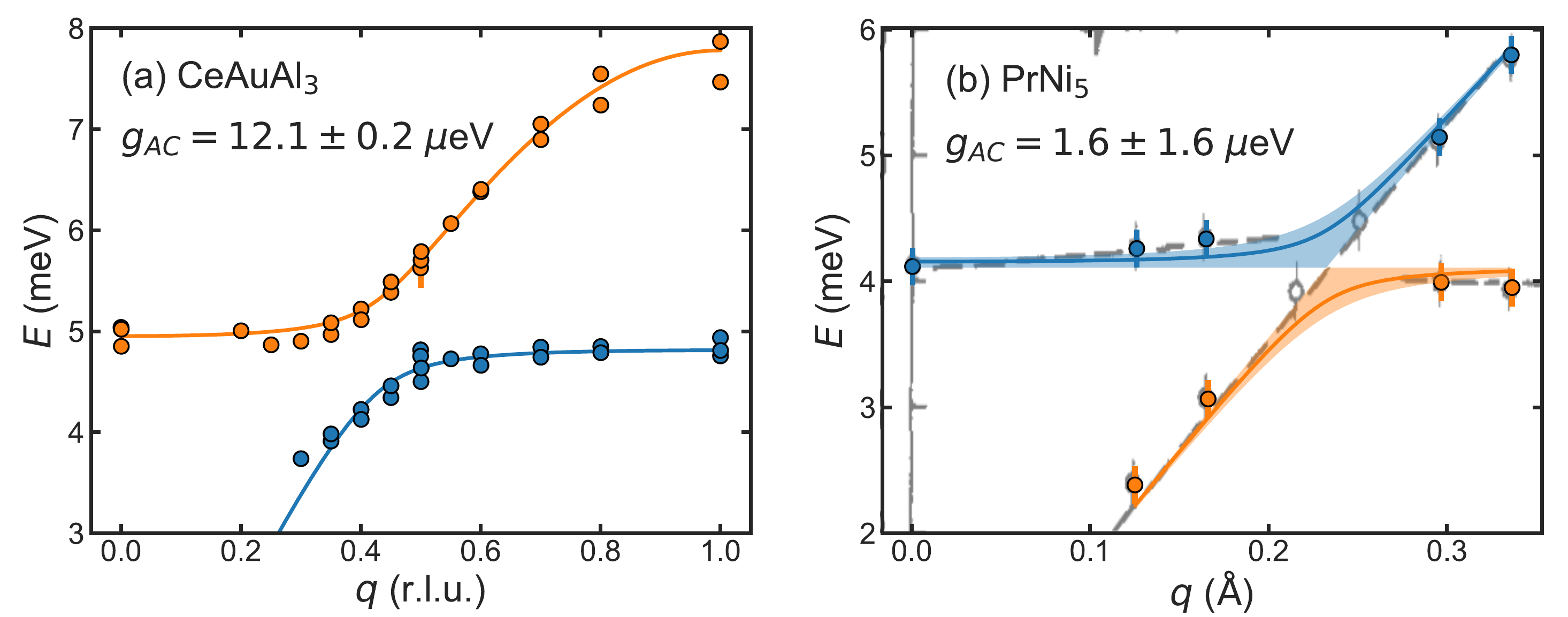}%
      \caption{\label{anticrossing-fit} (a) Anticrossing of the TA phonon branch and the crystal field excitation in CeAuAl$_3$. The solid lines are the result of the fit of the Eq.\,(\ref{eq-ac}) to the measured data.
      (b) Anticrossing in PrNi$_5$. Solid lines are result of the fit of the Eq.\, (\ref{eq-ac-aks}) to the data measured by Aksenov \cite{1983:aksenov:physbc}. Due to the lack of resolution, two nearest points to the crossing point were excluded from the fitting procedure, as there is not possible to distiguish peak splitting.
      The shaded area denotes possible solutions with a respect to the fitted error.
      Black underlying data in the background were taken from Ref.\,\cite{1983:aksenov:physbc}. On the basis of this fit we estimate the effective anticrossing coupling constant in PrNi$_5$ to be smaller than 4\,$\mu$eV. }
    \end{figure}

\noindent One may finally obtain the mixed mode dispersion as:
\begin{eqnarray}
  \label{eq-ac}
  \omega_{\mathbf{q}\pm}^2 = \frac{E_{\Gamma_6\Gamma_7^1}^2+\omega_0^2}{2}\mp\Big[\Big(\frac{E_{\Gamma_6\Gamma_7^1}^2-\omega_0^2}{2}\Big)^2
  +16 \alpha^2 E_{\Gamma_6\Gamma_7^1} \omega_0^2 g_{\rm AC} \Big]^{\frac{1}{2}},
  \end{eqnarray}
With equation (\ref{eq-ac}), the magnetoelastic constant may be inferred from the experimental data.
For the modelling of the acoustic phonon dispersion, the simplest possible model was used corresponding to a 1D chain:
\begin{equation}
  \label{eq-acoustic-phonon}
\omega_{q}= A \sqrt{1-\cos(|\bm{q}|)}  
\end{equation} 
  where $A$ is related to the amplitude of the branch at the border of the Brillouin zone. 
  At first data points were determined by Gaussian fits of the raw data, Fig. \ref{ac-all-fits}(a-h). This approach ignores intensities of the excitations and results in the set of data points shown in Fig. \ref{ac-all-fits}(i). The data were subsequently fitted with the model described by Eq.\,(\ref{eq-ac}) and (\ref{eq-acoustic-phonon}). The results of the fit are shown in Fig. \ref{anticrossing-fit} with all points measured also in different Brillouin zones (see Fig. 2 in the manuscript for color coding). The results of the fit may be summarized as follows: $A$ = 5.439(6) meV, $E_{\Gamma_6\Gamma_7^1}$ = 4.952(3) meV and $g_{\rm AC}\,=\,$12.1(2)\,$\mu$eV.

In order to compare our effective coupling constant with the results of Aksenov on PrNi$_5$ \cite{1983:aksenov:physbc}, we refitted their data. In particular, it appears that the dependence of the matrix elements $V_{mn}$ on the wavevector $q$ and a factor of 1/2 have not been taken into account. The expression we used instead to account for the coupled quadrupole-phonon excitation in PrNi$_5$ may be expressed as:
\begin{eqnarray}
  \label{eq-ac-aks}
  \omega_{\mathbf{q}\pm}^2 = \frac{E_{\Gamma_4\Gamma_{5A}}^2+\omega_0^2}{2}\mp\Big[\Big(\frac{E_{\Gamma_4\Gamma_{5A}}^2-\omega_0^2}{2}\Big)^2
  +\frac{9.82^2}{2} E_{\Gamma_4\Gamma_{5A}} \omega_0^2 g_{\rm AC} \Big]^{\frac{1}{2}},
  \end{eqnarray}
Refitting the data reported n PrNi$_5$ we obtain as an upper bound $g_{\rm AC}\,<\,$4\,$\mu$eV, which is much smaller than for \ceaual. An exact determination of the coupling constant is unfortunately due to resolution of the data as illustrated in Fig. \ref{anticrossing-fit}\,(b).
  
\section*{Vibronic bound state}
\label{VBS}
   
The total Hamiltonian may be written as
\begin{equation}
H_{total}=H_{CEF}+\hbar\omega_0(a_u^+a_u+1/2)-g(a_u+a_u^+)O_u,
\end{equation} 
where the first term is the tetragonal CEF Hamiltonian mentioned above, the second term is the phonon Hamiltonian ($H_{ph}$), and the third term is the c-ph coupling term. 
Here $\hbar\omega_0$ denotes the phonon energy, and $a_{\bm{u}}^+$ or $a_{\bm{u}}$ are phonon creation or annihilation operators, $\bm{u}$ is the phonon displacement, 
and $g$ is a magnetoelastic parameter proportional to the coupling between CEF and phonon excitations. $O_u$ is the CEF-phonon operator, with $u=\alpha, \gamma, \delta, 
\epsilon_1$ and $\epsilon_2$. According to ref.\cite{2012:adroja:prl}, the CEF-phonon operator should be of the form
\begin{equation}
O_{\gamma}=O_2^2=J_x^2-J_y^2=\frac{1}{2}(J_+^2+J_-^2).
\end{equation}
In the basis states of the CEF states, $O_\gamma$ is given by
\begin{equation}
O_\gamma=
\left(
  \begin{array}{cccccc}
    0 & 0 & O_\gamma^{13} & 0 & 0 & O_\gamma^{16} \\
    0 & 0 & 0 & O_\gamma^{13} & O_\gamma^{16} & 0 \\
    O_\gamma^{13} & 0 & 0 & 0 & 0 & 0 \\
    0 & O_\gamma^{13} & 0 & 0 & 0 & 0 \\
    0 & O_\gamma^{16} & 0 & 0 & 0 & 0 \\
    O_\gamma^{16} & 0 & 0 & 0 & 0 & 0 \\
  \end{array}
\right),
\end{equation}
with
\begin{equation}
O_\gamma^{13}\equiv\langle1|O_\gamma|3\rangle=-3\sqrt{2}\alpha+\sqrt{10}\beta,
\end{equation}
\begin{equation}
O_\gamma^{16}\equiv\langle1|O_\gamma|6\rangle=\sqrt{10}\alpha+3\sqrt{2}\beta.
\end{equation}
Then the oscillator strengths $Q_{\gamma}$ is given by
\begin{equation}
Q_{\gamma}=\left(
    \begin{array}{ccc}
      0 & 15.0928 & 40.9046 \\
      15.0928 & 0 & 0 \\
      40.9046 & 0 & 0 \\
    \end{array}
  \right)
\end{equation}
with $Q_{\alpha\beta}=\sum_{nm}{|\langle\Gamma_\alpha^m|O_\gamma|\Gamma_\beta^n\rangle|^2}$. 
We have $Q_\gamma^{\Gamma_6\Gamma_7^2}=Q_\gamma^{\Gamma_7^2\Gamma_6}=40.9046$ and $Q_\gamma^{\Gamma_6\Gamma_7^1}=Q_\gamma^{\Gamma_7^1\Gamma_6}=15.0928$. 
This shows that the phonon coupling is possible only to the inelastic $\Gamma_6\leftrightarrow\Gamma_7^1$ 
and $\Gamma_6\leftrightarrow\Gamma_7^2$ transitions.

The eigenstates of $H_{total}$ are calculated within the eight-dimensional subspace $|\psi_k\rangle, k=1,2,\cdots,8$ 
of zero- and one-phonon states $|\Gamma_6,0\rangle$, $|\Gamma_7^1,0\rangle$, $|\Gamma_7^2,0\rangle$ and $|\Gamma_6,1\rangle=a^+|\Gamma_6,0\rangle$ 
with unperturbed energies $0$, $E_{\Gamma_6\Gamma_7^1}$, $E_{\Gamma_6\Gamma_7^2}$, and $\hbar\omega_0$, respectively. 
The diagonalization of the Hamiltonian $H_{total}$ yields phonon-like states, which represent linear combinations of purely electronic $|\Gamma_7^2,0\rangle$ states 
and one-phonon states $|\Gamma_6,1\rangle$:
\begin{equation}
|\widetilde{\Gamma_7^2}\rangle=e|\Gamma_6,1\rangle+\sqrt{1-e^2}|\Gamma_7^2\rangle,
\end{equation}
This may be interpreted as bound states of a CEF excitation with phonons.

For a rough estimate of the coupling constant $g_{\rm VBS}$ we have assumed $\hbar\omega_0$ = 6.5 meV, just between $E_{\rm CF}$ and $E_{\rm VBS}$. This value is in agreement with an enhanced density of phonon states as determined in DFT calculations (see Fig. 2 in the manuscript). In the next step, the CEF Hamiltonian leading to unperturbed CEF transitions at $E_{\rm CF1} \approx \hbar\omega_0$ and $E_{\rm CF2} = 24.2$\,meV was guessed. In turn, the coupling constant was fitted to the measured energy transitions at $E_{\rm CF1}$, $E_{\rm VBS}$ and $E_{\rm CF2}$ resulting in $g_{\rm VBS} \approx 400\,{\rm \mu eV}$. This approximation is based on a simple model assuming non-dispersive phonon states. For the correct treatment of the vibronic coupling in \cecual one needs to involve a more complex theoretical apparatus. 

\section*{Additional figures and resources}

\begin{figure}[ht]
\centering
  \includegraphics[width=0.4\linewidth]{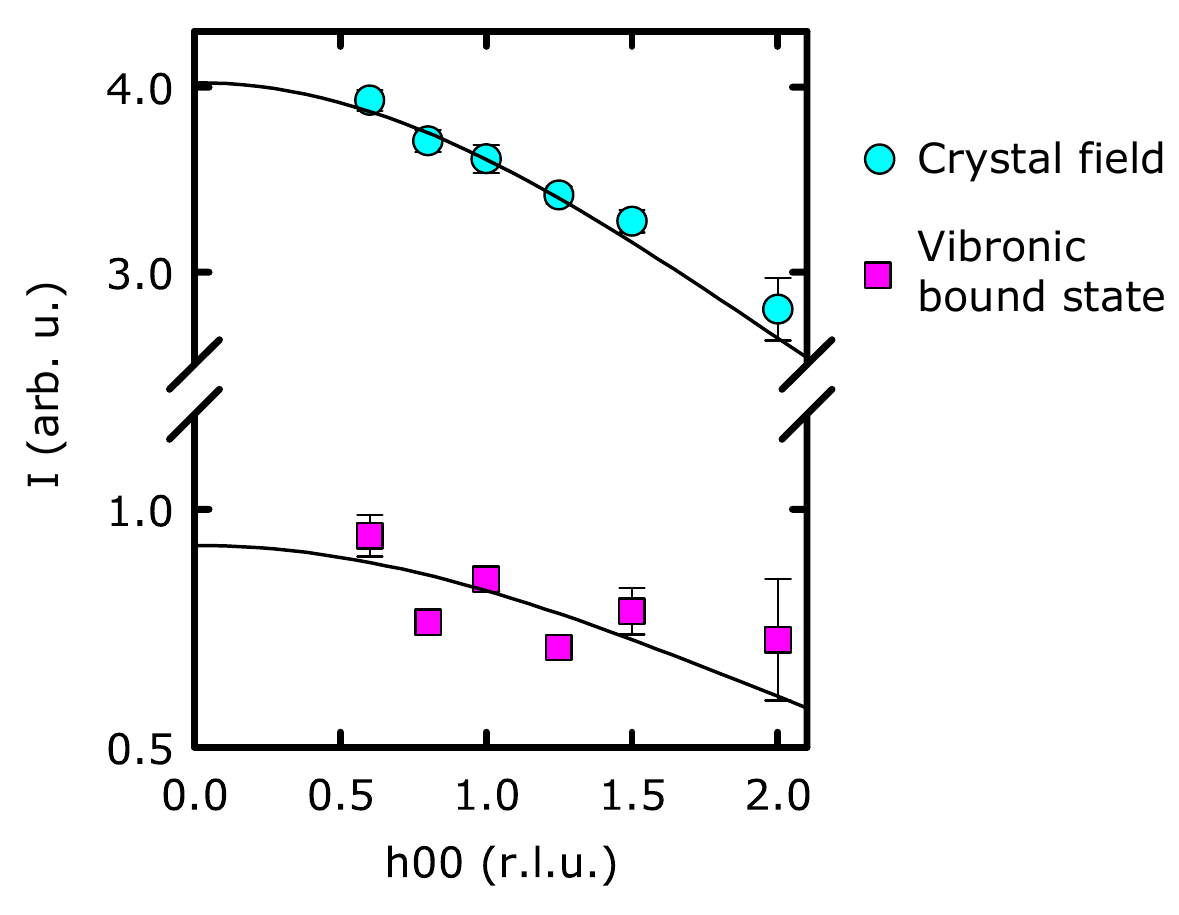}
  \caption{\label{h00-qdep} $Q$-dependence of the integrated intensities of the crystal field and vibronic bound state along the ($h$, 0, 0) direction. 
  The black line denotes a fit to the form factor of the Ce$^{3+}$ ion. 
  The data at the last point, (2, 0, 0), were measured for a different sample orientation and with lower statistics, accounting for the larger error bars.  }
\end{figure}

In the following additional figures are presented to illustrate further the phenomena described in the main text. Fig. \ref{h00-qdep} shows the momentum dependence of the crystal field and vibronic excitations, which decreases with increasing momentum following closely the form factor of the Ce$^{3+}$ ion (solid line).

\begin{figure}[ht]
\centering
 \includegraphics[width=0.3\linewidth]{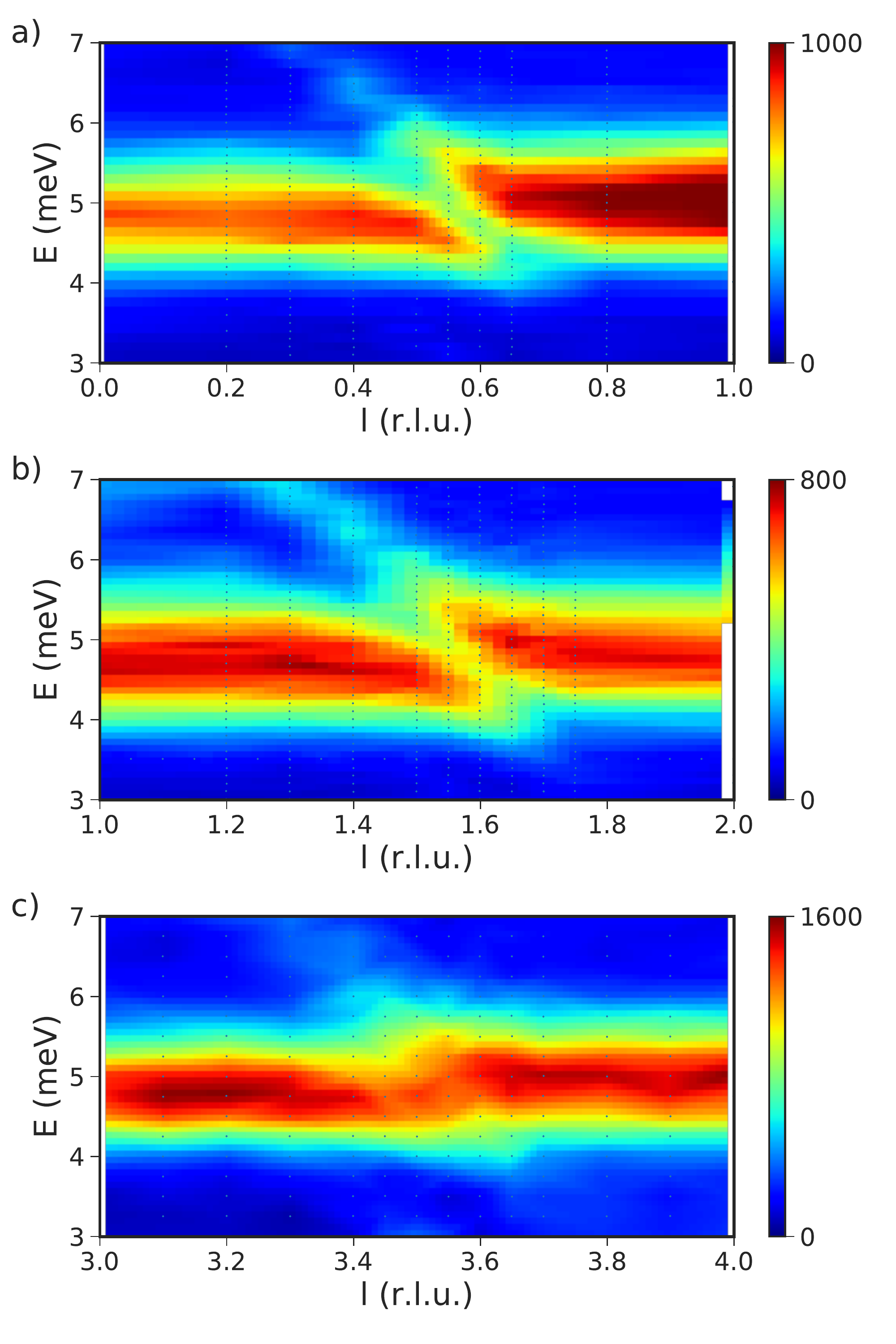}
 \caption{\label{allAC} Intensity maps measured on PUMA instrument in the a) 10L, b) 20L and c) 11L region.}
\end{figure}

The anticrossing described in the main text was observed for different positions in $Q$-space. The effect was not always evident due to the tilt of resolution elipsoid. In order to illustrate this, data across the anticrossing are shown for three different Brillouin zones in Fig. \ref{allAC}. Fig. \ref{allAC}(a) was also shown in the main text. It displays the anticrossing nearby a weak reflection (101), notice enhanced intensity of the CF mode between anticrossing and middle of the Brillouin zone 101. This effect may be caused by an underlying softened optical phonon.  Polarized neutron scattering experiments will be needed to resolve this issue. Further, Fig. \ref{allAC}(b) shows the anticrossing nearby the strong elastic reflection at (202). Here the CEF intensity is constant on both sides of the anticrossing and the phonon branch across the anticrossing is enhanced. The anticrossing effect is, finally, barely visible nearby the (114) $\Gamma$-point due to the instrumental resolution as shown in Fig. \ref{allAC}(c).

\begin{figure}[ht]
\centering
   \includegraphics[width=0.6\linewidth]{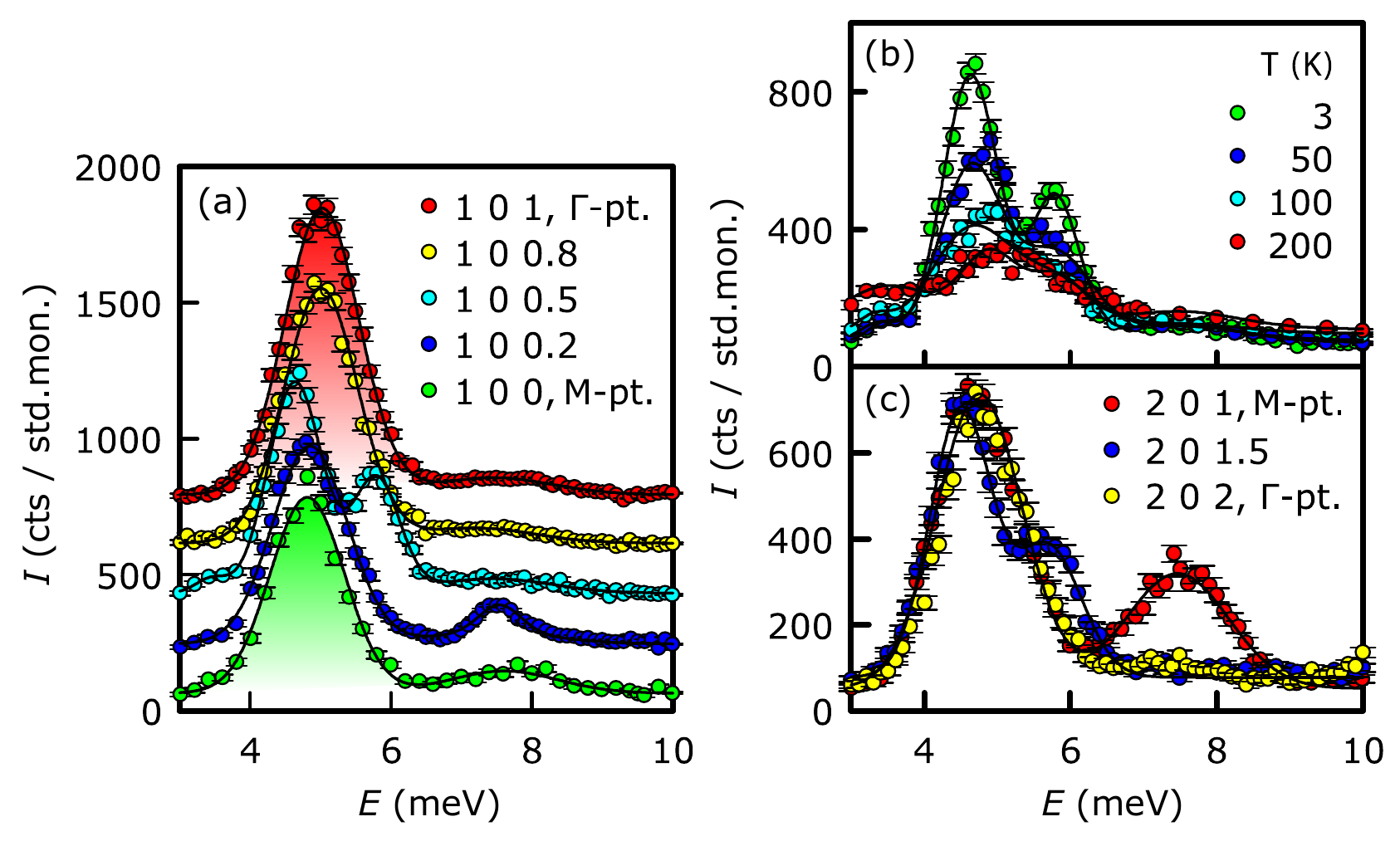}
   \caption{\label{rest} Constant $Q$ scans normalized to the standard monitor.
   Black lines are results of the fit of Gaussians. 
    (a) Comparison of the width and intensities of the anticrossing in the (101) Brillouin zone.
   Note the unusual enhancement of the CF intensity which decreases with lower $Q$ (from red to green). 
   Data are shifted vertically by 200 cts for clarity.
   (b) Temperature dependence of the anticrossing between phonon and crystal field at the (1 0 0.5) point. 
   See that the whole intensity scales down as expected from the crystal field level. 
   Lines are Gaussian fits to the data.
   (c)  Comparison of the width and intensities of the anticrossing in the (202) Brillouin zone. }
\end{figure}

Detailed energy scans through different $Q$-points of the anticrossing in the (101) Brillouin zone is shown in Fig. \ref{rest}(a) and its temperature dependence is in Fig. \ref{rest}(b). The enhanced intensity in the middle of the Brillouin zone ($\Gamma$-point, red area) as compared to the M-point (green area) was already discussed in the main text. This effect is not visible for the (202) $\Gamma$-point, cf. Fig. \ref{rest}(c).

\newpage
\bibliography{bibliography}